\documentclass[12pt]{article}
\usepackage{graphicx}
\newcommand{\CP}{{\cal P}}
\newcommand{\CQ}{{\cal Q}}
\newcommand{\CR}{{\cal R}}
\newcommand{\CG}{{\cal G}}
\newcommand{\CV}{{\cal V}}
\newcommand{\CD}{{\cal D}}
\makeatletter
	
		\@addtoreset{equation}{section}
\makeatother

\makeatletter
\@addtoreset{equation}{section}
\makeatother

\begin{document} %
\begin{flushright}
{KOBE-TH-05-03}\\
{OU-HET-530}\\
\end{flushright}
\vspace{5mm}
\vspace*{6mm}
\begin{center}
{\Large \bf Extended supersymmetry and its reduction \\ on a circle with point singularities
}
\vspace{10mm} \\
Tomoaki Nagasawa\footnote{e-mail:
tnagasaw@kobe-u.ac.jp (T. Nagasawa)},
Makoto Sakamoto\footnote{e-mail:
dragon@kobe-u.ac.jp (M. Sakamoto)},
Kazunori Takenaga\footnote{e-mail:
takenaga@het.phys.sci.osaka-u.ac.jp (K. Takenaga)}
\vspace*{10mm} \\
{\small \it
${}^1 $Graduate School of Science and Technology, Kobe University,
Rokkodai, Nada, \\ Kobe 657-8501, Japan
\\
${}^2 $Department of Physics, Kobe University,
Rokkodai, Nada, Kobe 657-8501, Japan
\\
${}^3 $Department of Physics, Osaka University, 
Toyonaka 560-0043, Japan
}
\vspace{15mm}
\\
{\bf Abstract}
\vspace{4mm}
\\
\begin{minipage}[t]{130mm}
\baselineskip 6mm 

We investigate $N$-extended supersymmetry in one-dimensional quantum mechanics on a circle with point singularities.
For any integer $n$, $N=2n+1$ supercharges are explicitly constructed in terms of 
discrete transformations, 
and a class of singularities  
compatible with supersymmetry is clarified.
In our formulation, the 
supersymmetry can be reduced to $M$-extended supersymmetry for any integer $M<N$. 
The degeneracy of the spectrum and spontaneous supersymmetry breaking are also studied. 

\end{minipage}
\end{center}
\vspace{8mm}
\baselineskip 6mm
\section{Introduction}

A point singularity in one dimensional quantum mechanics may be considered, in general, as a localized limit of a finite range potential and is parametrized by the group $U(2)$\cite{U(2)1,U(2)2,U(2)3}, and the parameters characterize connection conditions between a wavefunction and its derivative at the singularity.
The varieties of the connection conditions lead to various interesting phenomena, such as duality\cite{duality1,duality2}, the Berry phase\cite{Berry1,Berry2}, scale anomaly\cite{Anomaly} and supersymmetry\cite{susy1,susy2,susy3,susy4}.

\nopagebreak[4]
$N=1$ or $N=2$ supersymmetry in the model of a free particle on a line {\bf R} or an interval $[-l, l]$ with a point singularity was discussed in Ref. \cite{susy1}.
In Ref. \cite{susy3}, this work was extended to $N=4$ supersymmetry in the model on a pair of lines or intervals each having a point singularity.
In Ref. \cite{susy2}, $N=2$ supersymmetric model with a superpotential was constructed on a circle with two point singularities.
The supercharges are represented in terms of a set of discrete transformation $\{ \CP_1, \CQ_1, \CR_1\}$, which forms an $su(2)$ algebra of spin $1/2$.
In Ref. \cite{susy4}, this work was extended to $N=2n$ supersymmetry by putting $2^n$ number of point singularities on a circle, and $N=2n$ supercharges are explicitly constructed in terms of $n$ sets of  discrete transformations $\{\CP_1, \CQ_1, \CR_1\}, \cdots, \{\CP_n,\CQ_n,\CR_n\}$ on the circle.

We would like to emphasize that the study on the supersymmetry in one dimensional quantum mechanics with point singularities has a physical application to higher dimensional gauge theories.
It has been shown that a quantum-mechanical supersymmetry is hidden in any gauge invariant theories with extra dimensions\cite{Shaposhnikov-Tinyakov, Nagasawa-Sakamoto,Lim-nagasawa-sakamoto-sonoda}.
In Ref. \cite{Nagasawa-Sakamoto}, the hierarchy problem at tree level has been solved in a scenario of gauge theories with compact extra dimensions with boundaries\cite{Higgsless}.
Then, the hidden quantum-mechanical supersymmetric structure as well as the choice of boundary conditions have been found to be crucial, and the analyses in Ref. \cite{susy2,susy4} have turned out to give a powerful tool to derive all possible sets of boundary conditions.


In this paper, we study $N=2n+1$ supersymmetry on a circle with $2^n$ point singularities, and give a full detail of the analysis.
The $N=2n+1$ supercharges are explicitly constructed in terms of $n+1$ sets of discrete transformations $$\{ \CP_1, \CQ_1, \CR_1 \}, \cdots ,\{ \CP_n,\CQ_n, \CR_n \}, \{\CP_{n+1}, \CQ_{n+1}, \CR_{n+1} \}.$$ 
Since the model contains point singularities, we need to impose appropriate connection conditions there.
We succeed to clarify a possible set of 
connection conditions compatible with the $N=2n+1$ supercharges. 
Thus, the $N=2n+1$ supersymmetry can be realized under the connection conditions. 
In our formulation, we can remove some of the $N=2n+1$ supercharges from a class of physical observables 
by relaxing the connection conditions, so that 
the $N=2n+1$ supersymmetry can be reduced to $M$-extended supersymmetry for any integer $M<N$.
This implies that for any fixed $M$ we have a wide variety of $M$-extended supersymmetric models.
We provide a general discussion about reduction of the supersymmetry and construct $N=2n$ supersymmetric models 
as a result of the reduction of the original $N=2n+1$ supersymmetry. 
We also investigate the degeneracy of the spectrum, in particular, vacuum states with vanishing energy. 

The plan of this paper is as follows.
In Section 2, we introduce $n+1$ sets of discrete transformations on a circle. 
In Section 3, we construct $N=2n+1$ supercharges in terms of these discrete transformations and examine
connection conditions compatible with all the $2n+1$ supercharges.
We also investigate 
the degeneracy of the spectrum. 
 In Section 4, we discuss reduction of the supersymmetry and study some $N=2n$ supersymmetric models as examples.
Section 5 is devoted to summary. 
\section{Discrete transformations}

We consider the model which is one-dimensional quantum mechanics on a circle $S^1\ (-l <x \le l)$ with $2^n$ point singularities placed at 
\begin{equation}	
x= l_s \equiv 
	\left(
		1- \frac{s}{2^{n-1}} 
	\right)l
	\qquad {\rm for} \ s=0,1,\cdots ,2^n-1.
\end{equation}
This paper deals with the model 
in which the wavefunction and its derivative are continuous everywhere except for the point singularities.
These point singularities are placed at regular intervals on the circle (Fig. 1).
\begin{figure}[t]
\begin{center}
	\includegraphics{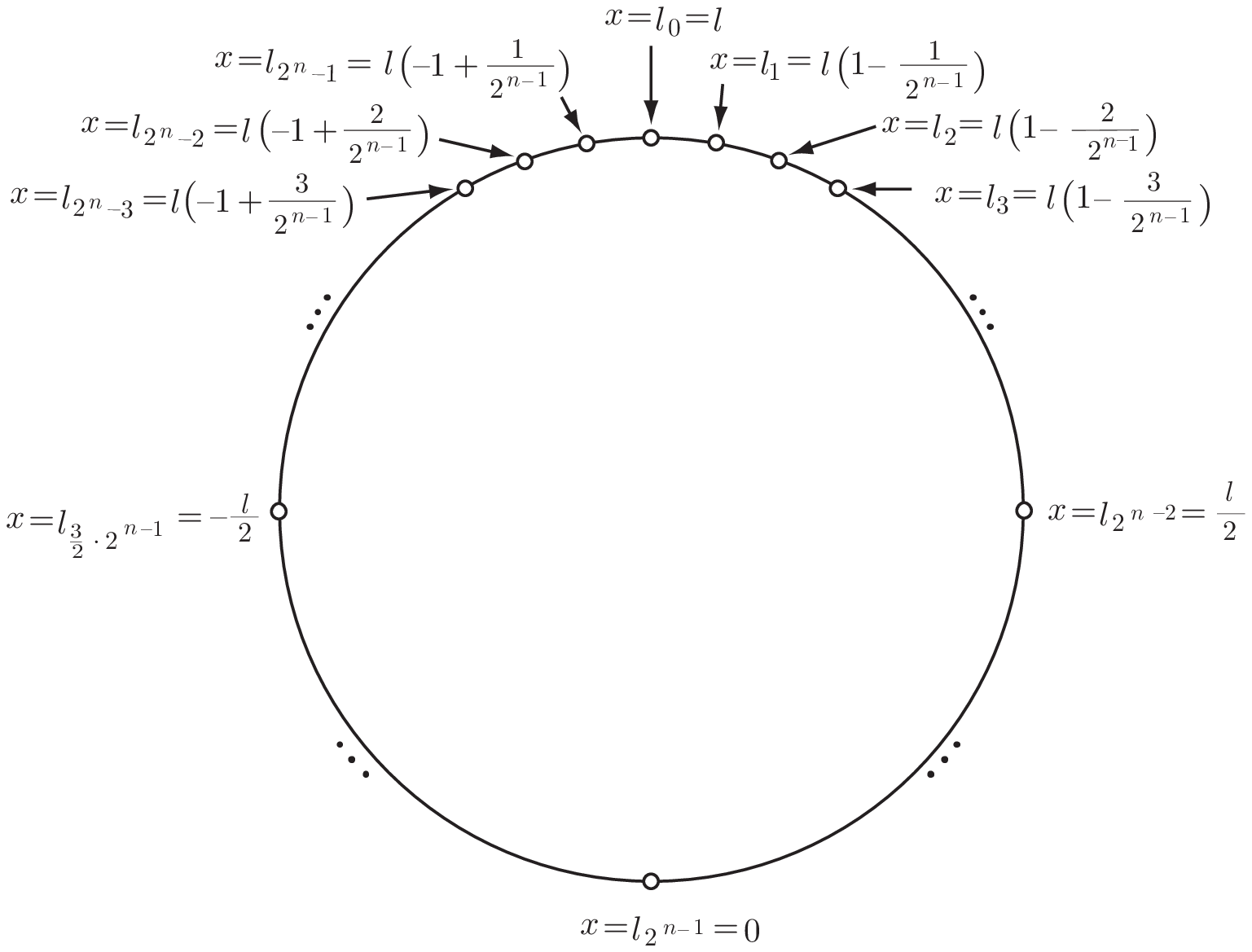}
\end{center}
\caption{The positions of the singularities on a circle $S^1(-l <x \le l)$
}
\end{figure}
We define discrete transformations on the circle as
\begin{eqnarray}
	&&(\CP_k \varphi )(x)=\sum_{s=1}^{2^{k-1}} 
	\Theta
		\left(
			x-\left(1-\frac{s}{2^{k-2}}
				\right)
			l
		\right)
	\Theta
		\left(
			\left(1-\frac{s-1}{2^{k-2}}
				\right)
			l-x
		\right) 
	\nonumber \\
	&&\qquad \qquad \qquad \qquad \qquad \qquad \qquad \qquad \qquad \times \varphi
		\left(
			-x+\left(2-\frac{2s-1}{2^{k-1}}
				\right)
			l
		\right),
	\\
	&&(\CR_k \varphi )(x)=\sum_{s=1}^{2^{k-1}} 
	(-1)^s
	\left[
	-\Theta
		\left(
			x-\left(1-\frac{s-1/2}{2^{k-2}}
				\right)
			l
		\right)
	\Theta
		\left(
			\left(1-\frac{s-1}{2^{k-2}}
				\right)
			l-x
		\right) 
	\right.
	\nonumber \\
	&&
	\qquad \qquad \qquad \qquad \qquad 
	+
	\left.
	\Theta
		\left(
			x-\left(1-\frac{s}{2^{k-2}}
				\right)
			l
		\right)
	\Theta
		\left(
			\left(1-\frac{s-1/2}{2^{k-2}}
				\right)
			l-x
		\right) 
	\right]
	\varphi(x) , 
	\nonumber \\
	\\
	&&(\CQ_k \varphi)(x) \equiv -i \left( \CR_k \CP_k \varphi \right)(x) 
\end{eqnarray}
for $k=1,2,\cdots,n+1$.
Here $\Theta(x)$ is the Heaviside step function defined by
\begin{eqnarray}
	\Theta(x) =\left\{ 
			\begin{array}{lcl}
			      1 & {\rm for} & 0<x<l, \\
				 0 & {\rm  for}& -l <x< 0.
			\end{array}
		\right. 
\end{eqnarray}
The $\CP_k \ (k=1,2,\cdots, n+1)$ is a kind of the parity transformation.
The $\CP_1$ is just the familiar parity transformation, $\left(\CP_1 \varphi \right)(x) =\varphi(-x)$.
The $\CR_k\ (k=1,2,\cdots,n+1)$ is a kind of the half-reflection transformation.
The action of $\CP_k,\CR_k$ $(k=1,2,3)$ is schematically depicted in Fig. 2- Fig. 4.
In each figure, the upper figures mean the geometrical meaning of $\CP_k,\CR_k$ $(k=1,2,3)$. 
In the lower figures, the dashed line denotes the original function $\varphi(x)$, and the solid line denotes $\left(\CP_k \varphi \right)(x), \left( \CR_k \varphi\right) (x)$ $(k=1,2,3)$.
\begin{figure}[h]
\begin{center}
	\includegraphics{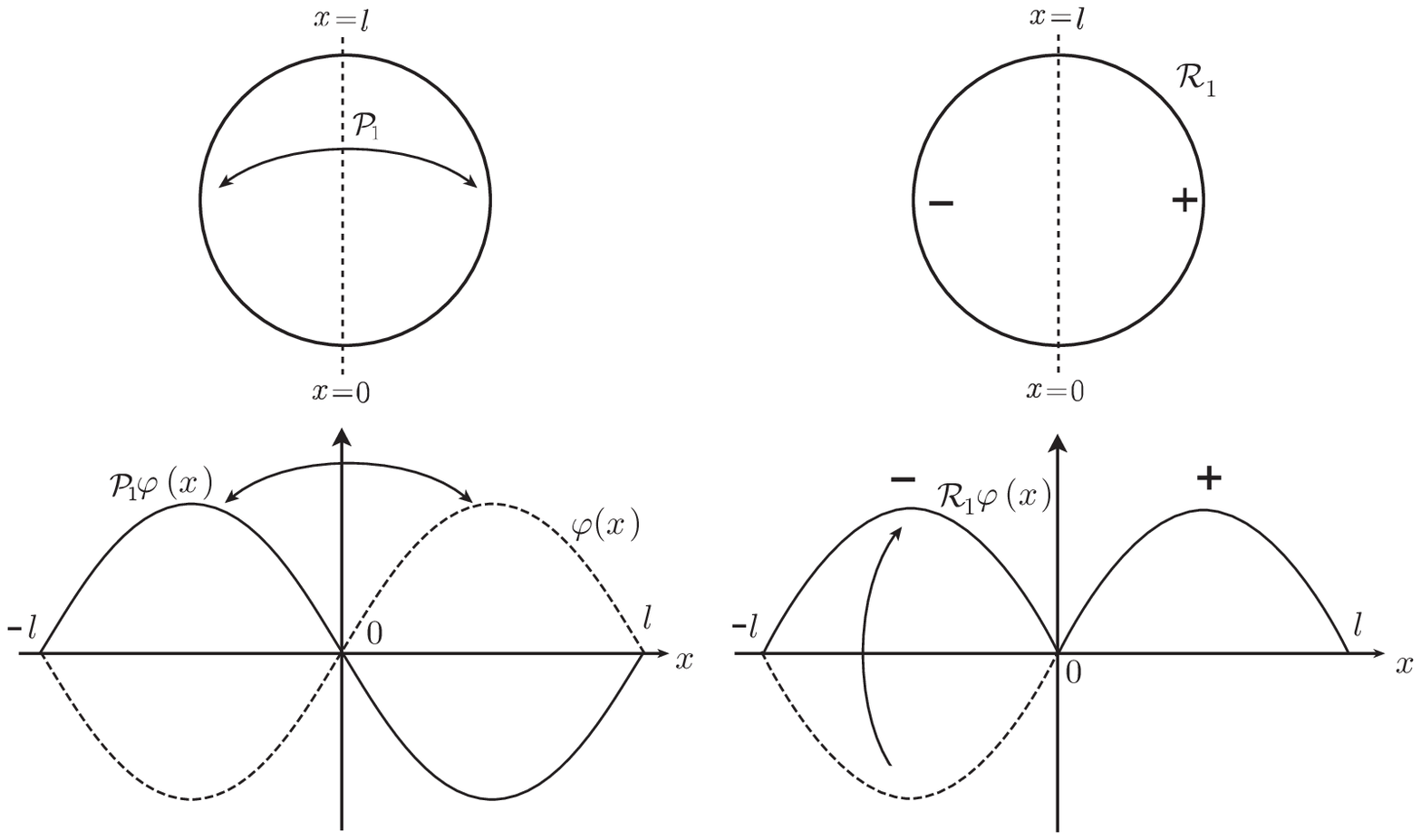}
		\caption{The action of 
$\CP_1, \CR_1$
}
\end{center}
\end{figure}
\begin{figure}[htbp]
\begin{center}
	\includegraphics{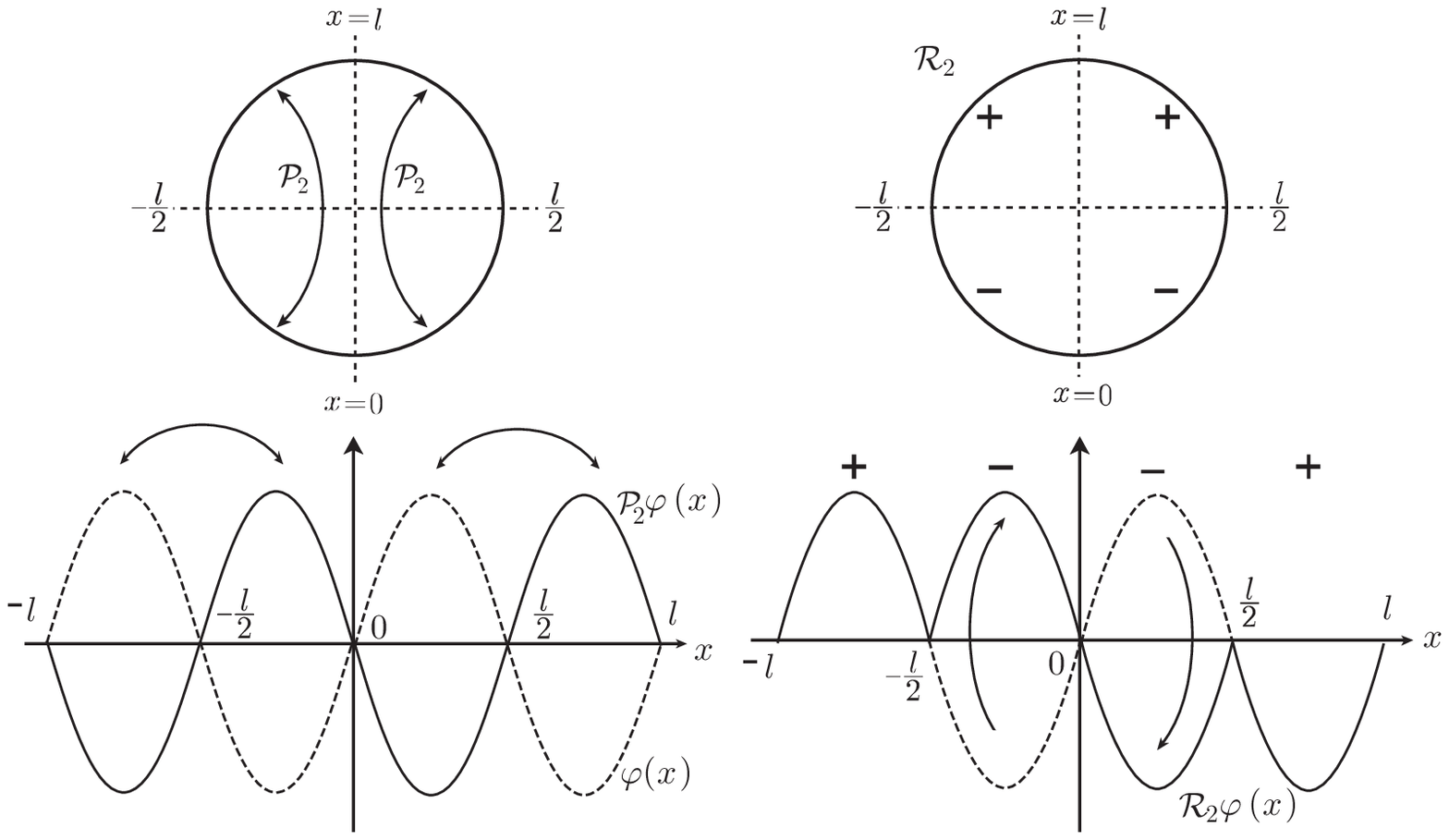}
\end{center}
\caption{ The action of 
$\CP_2, \CR_2$ }
\end{figure}
\begin{figure}[htbp]
\begin{center}
	\includegraphics{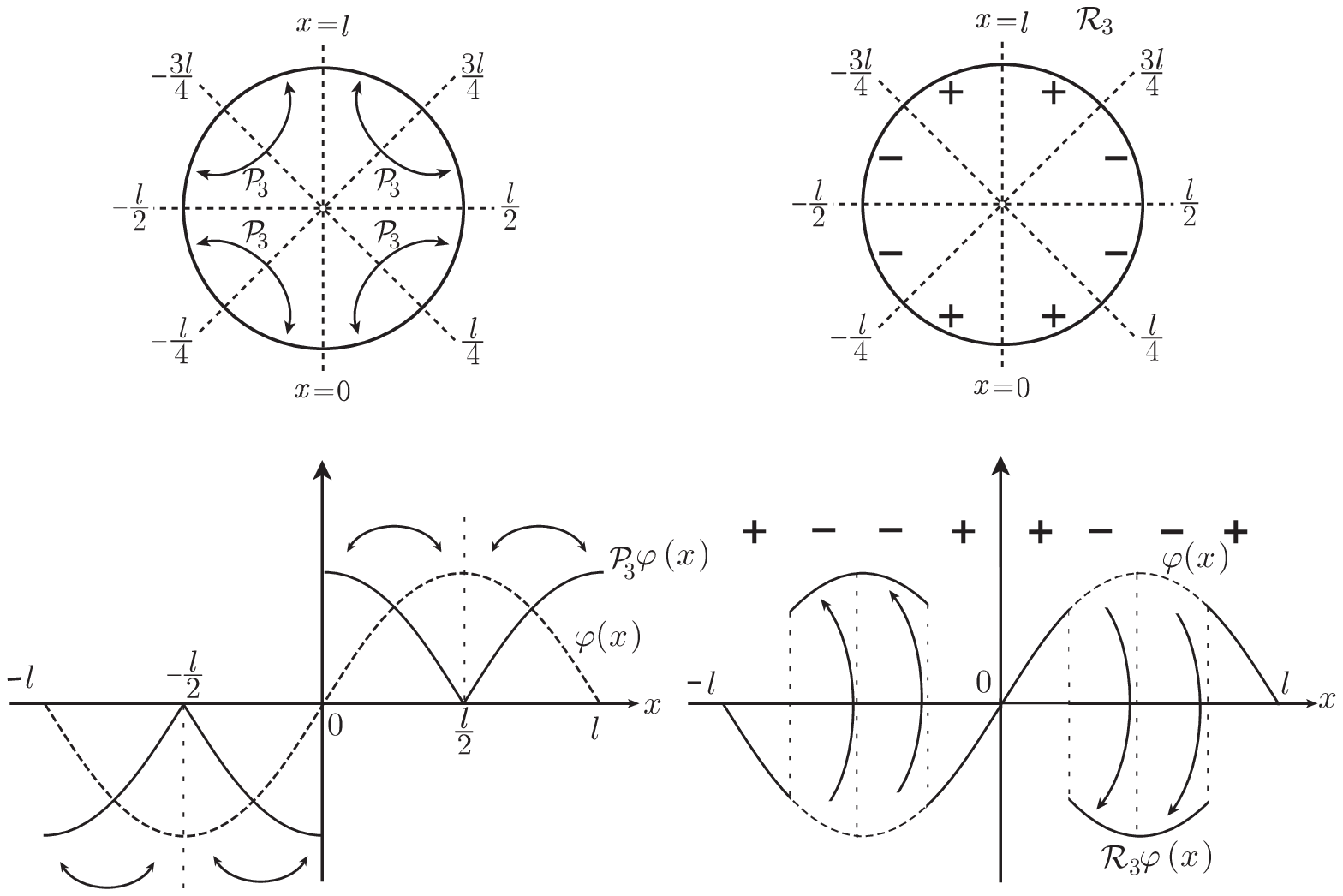}
\end{center}
\caption{The action of 
$\CP_3, \CR_3$
}
\end{figure}

The ${\CP_k ,\CQ_k ,\CR_k}$ ($k=1,2,\cdots ,n$) and $\CP_{n+1}$ produce no new singularities except for the original ones.
Note that $\CQ_{n+1}$ and $\CR_{n+1}$ make wavefunction and/or its derivative discontinuous at $x\ne l_s$ $(s=1,2,\cdots,2^n-1)$.
For instance, if we consider the case of $n=1$, 
there are two singularities located at $x=0,l$.
As seen in fig. 2, a set of discrete transformation $\{ \CP_1, \CQ_1, \CR_1 \}$ does not make wavefunction and its derivative discontinuous at $x \ne 0,l$. 
We introduce a new set of discrete transformation $\{ \CP_2, \CQ_2, \CR_2 \}$.
Although the $\CP_2$ produces no new discontinuity for wavefunction and its derivative at $x\ne 0,l$,  
the $\CQ_2$ and $\CR_2$, however,  produce discontinuity for wavefunction and/or its derivative 
at the different points of the original singularities, $x=\pm\frac{l}{2}$ (see fig. 3).

A crucial observation is that each set $\{ \CP_k ,\CQ_k, \CR_k \}\ (k=1,2,\cdots,n+1)$ forms an $su(2)$ algebra of spin $1/2$, 
\begin{eqnarray}
&&\CP_k \CQ_k =-\CQ_k \CP_k =i\CR_k, \quad 
\CQ_k \CR_k =-\CR_k \CQ_k =i\CP_k, \quad 
\CR_k \CP_k =-\CP_k \CR_k =i\CQ_k, \nonumber \\
&& 
\left( \CP_k \right)^2 =\left( \CQ_k \right)^2 =\left( \CR_k \right)^2 =1  \label{su(2)}
\end{eqnarray}
and that ${\cal O}_k = \{ \CP_k ,\CQ_k, \CR_k \}$ and ${\cal O}_{k'} =\{ \CP_{k'} ,\CQ_{k'}, \CR_{k'} \}$ commute with each other if $k\ne k'$, 
\begin{equation}
	[{\cal O}_k , {\cal O}_{k'}]=0 \qquad {\rm for} \ k\ne k'.
\end{equation}

For later use, let us introduce new sets of $su(2)$ generators $\{ \CG_{\CP_{k}}, \CG_{\CQ_{k}}, \CG_{\CR_k} \}$ $(k=1,2,\cdots,n)$ as 
\begin{eqnarray}
	\CG _{\CP_k} = \CV^{\dagger} \CP_k \CV, \quad 
	\CG _{\CQ_k} = \CV^{\dagger} \CQ_k \CV, \quad 
	\CG _{\CR_k} = \CV^{\dagger} \CR_k \CV,  
	\qquad 
	\ k=1,2,\cdots ,n, \label{singular}
\end{eqnarray}
where
\begin{eqnarray}
	\CV \equiv \CV_1\CV_2 \cdots \CV_n , \quad 
	\CV_k =e^{i \vec{v}_k \cdot \vec{\CP_{k}}} \in SU(2). \label{SINGULAR2}
\end{eqnarray}
Here, $\vec{\CP}_k\ (\vec{v}_k)$ is an abbreviation of $\vec{\CP}_k=(\CP_k, \CQ_k,\CR_k )\ (\vec{v}_k=( v_{\CP_k}, v_{\CQ_k}, v_{\CR_k}))$, and $( v_{\CP_k}, v_{\CQ_k}, v_{\CR_k})$ are real parameters.  
The new $su(2)$ generators $\{ \CG_{\CP_k} ,\CG_{\CQ_k},\CG_{\CR_k} \}$ have to be linearly related to $\vec{\CP}_k$ as 
\begin{eqnarray}
	\CG_{\CP_k}= \vec{e}_{\CP_k} \cdot \vec{\CP}_k, \quad 
	\CG_{\CQ_k}= \vec{e}_{\CQ_k} \cdot \vec{\CP}_k, \quad 
	\CG_{\CR_k}= \vec{e}_{\CR_k} \cdot \vec{\CP}_k, \qquad 
	\quad  k=1,2,\cdots,n,
\end{eqnarray}
where $\{ \vec{e}_{\CP_k} , \vec{e}_{\CQ_k}, \vec{e}_{\CR_k} \}$ are three-dimensional orthogonal unit vectors.
\section{\boldmath{$N=2n+1$} supersymmetry}

\subsection{\boldmath{$N=2n+1$} supercharges}

Equipped with the discrete transformations given in the previous section, we construct $N=2n+1$ supercharges in terms of the $n+1$ sets of the discrete transformations.
There are two types of $N=2n+1$ supercharges (type A and type B), 
\begin{itemize}
\item Type A
\end{itemize}
\begin{eqnarray}
	&&Q_a^A=\frac{i}{2} \Gamma_a \CD^{A}=\frac{i}{2} \bar{\CD}^{A} \Gamma_a, \quad
	a=1,2,\cdots, 2n+1, \label{Q_A}
\end{eqnarray}
\begin{itemize}
\item Type B
\end{itemize}
\begin{eqnarray}
	&&Q_a^B=\frac{i}{2} \Gamma_a \CD^{B}=\frac{i}{2} \bar{\CD}^{B} \Gamma_a, \quad 
	a=1,2,\cdots, 2n+1, \label{Q_B}
\end{eqnarray}
where
\begin{eqnarray}
	\CD^A &=& \left( \CR_1 \cdots \CR_{n+1} \frac{d}{dx} \right)+\CG_{\CR_1}\cdots \CG_{\CR_n} \left( \CR_1 \cdots \CR_{n+1} W'(x) \right), \\
	\bar{\CD}^A&=&\left( \CR_1 \cdots \CR_{n+1} \frac{d}{dx} \right)-\CG_{\CR_1}\cdots \CG_{\CR_n} \left( \CR_1 \cdots \CR_{n+1} W'(x) \right),  \\
	\CD^B &=& \left( \CR_1 \cdots \CR_{n+1} \frac{d}{dx} \right)-\CP_{n+1} \left( \CR_1 \cdots \CR_{n+1} V'(x) \right), \\
	\bar{\CD}^B&=&\left( \CR_1 \cdots \CR_{n+1} \frac{d}{dx} \right)+\CP_{n+1} \left( \CR_1 \cdots \CR_{n+1} V'(x) \right) 
\end{eqnarray}
and
\begin{eqnarray}
\begin{array}{lcl}
	\Gamma_{2k-1} &=&\CG_{\CR_1} \CG_{\CR_2} \cdots \CG_{\CR_{k-1}} \CG_{\CP_k} \CR_{n+1}, \\
	\Gamma_{2k} &=&\CG_{\CR_1} \CG_{\CR_2} \cdots \CG_{\CR_{k-1}} \CG_{\CQ_k} \CR_{n+1}, \quad 
	 k=1,2,\cdots,n, \\
	\Gamma_{2n+1} &=& \CQ_{n+1}.
\end{array}
\end{eqnarray}
Here, $W'(x)=\frac{d}{dx} W(x), V'(x)=\frac{d}{dx}V(x)$, and $W(x),V(x)$ are called superpotentials.
We note that $\Gamma_a , \CD^{A(B)}, \bar{\CD}^{A(B)}$ in the supercharges contain $\CR_{n+1},\CQ_{n+1}$ 
which produce new singularities except for the original singular point $x=l_s (s=1,2,\cdots, 2^n-1)$.
The combinations of these operators in the supercharges, however, make the $\CQ_{n+1}$ and $\CR_{n+1}$ vanished due to the $su(2)$ algebra. 

The functions $W'(x), V'(x)$ are continuous and finite valued functions at intervals between singularities 
and are allowed to have discontinuities at  
$x=l_s\ (s=0,1,\cdots, 2^n-1)$.
In order to construct the $N=2n+1$ superalgebra, the functions turn out to be required to 
\begin{eqnarray}
\CP_k W'(x) &=& -W'(x) \CP_k, \qquad  k=1,2,\cdots,n, \label{3-8}\\
\CP_{n+1} W'(x) &=&  W'(x) \CP_{n+1} \label{2n+1W}
\end{eqnarray}
and 
\begin{eqnarray}
\CP_i V'(x) &=& - V'(x) \CP_i,  \qquad  i=1,2,\cdots,n+1. \label{3-10}
\end{eqnarray}
Noting that 
$\CR_1 \cdots \CR_{n+1} \frac{d}{dx}$, $\CR_1 \cdots \CR_{n+1} W'(x)$ and $\CR_1\cdots \CR_{n+1} V'(x)$ commute with $\{ \CP_k ,\CQ_k, \CR_k\}$ $(k=1,2,\cdots,n+1)$,
we have the relations 
\begin{eqnarray}
	\{ \Gamma_a ,\Gamma_b \} &=&2\delta_{ab} ,\\
	 \Gamma_a {\CD}^{A(B)} &=& \bar{\CD}^{A(B)} \Gamma_a,  \qquad  a,b=1,2,\cdots,2n+1. \label{GD}
\end{eqnarray}
It follows that the supercharges form the $N=2n+1$ superalgebra;
\begin{itemize}
	\item Type A
\end{itemize}
\begin{eqnarray}
	 \{Q_a^A ,Q_b^A \} =H^{A} \delta_{ab}, \quad  
	 a,b=1,2,\cdots 2n+1, \label{SuperA1}
\end{eqnarray}
\begin{eqnarray}
	H^A &=&-\frac{1}{2} \bar{\CD}^A \CD^A =\frac{1}{2} \left[
		-\frac{d^2}{dx^2} -\CG_{\CR_1}\cdots \CG_{\CR_n}W''(x)+\left(W'(x)\right)^2 \right], \label{3-14}
\end{eqnarray}
\begin{itemize}
	\item Type B
\end{itemize}
\begin{eqnarray}
	 \{Q_a^B ,Q_b^B \} =H^{B} \delta_{ab}, \quad  
	 a,b=1,2,\cdots 2n+1,
\end{eqnarray}
\begin{eqnarray}
	H^B &=&-\frac{1}{2} \bar{\CD}^B \CD^B=\frac{1}{2} \left[
		-\frac{d^2}{dx^2} +\CP_{n+1} V''(x) +\left(V'(x)\right)^2 \right], \label{SuperA2}
\end{eqnarray}
where $H^A$ and $H^B$ are the Hamiltonian in each model.
\subsection{Connection conditions compatible with supersymmetry}

In this section, we clarify the connection conditions compatible with the $N=2n+1$ supersymmetry.
Since the model contains the point singularities, we need to impose appropriate connection conditions there, namely, 
our model is specified not only by the Hamiltonian but also by the connection conditions. 
The functional space in our model is required to be squared integrable and to be spanned by eigenfunctions of the Hamiltonian with connection conditions that  
make the Hamiltonian hermitian.

In order to obtain the appropriate connection conditions, let us introduce a $2^{n+1}$ -dimensional boundary vector $\Phi_{\varphi}$ that consists of boundary values of a wavefunction $\varphi(x)$ in the vicinity of the singularities, $\varphi(l_s \pm \epsilon)$ for $s=0,1,\cdots,2^n-1$ with an infinitesimal positive constant $\epsilon$. 
For later convenience, we arrange $\varphi(l_s \pm \epsilon)$ in such a way that $\Phi_{\varphi}$ satisfies the relations
\begin{eqnarray}
	&&\Phi_{{\cal P}_k \varphi} = 
		(\overbrace{I_2 \otimes \cdots \otimes I_2 \otimes 
		\sigma_1}^{k} \otimes I_2 \otimes \cdots \otimes I_2) 
		\Phi_{\varphi}, \label{B1} \\
	&&\Phi_{{\cal Q}_k \varphi} = 
		(I_2 \otimes \cdots \otimes I_2 \otimes \sigma_2 
		\otimes I_2 \otimes \cdots \otimes I_2) \Phi_{\varphi}, \label{B2} \\
	&&\Phi_{{\cal R}_k \varphi} = 
		(\underbrace{ I_2 \otimes \cdots \otimes I_2 \otimes 
		\sigma_3\otimes I_2 \otimes \cdots \otimes I_2}_{n+1}) 
		\Phi_{\varphi}, \label{B3}
\end{eqnarray}
where $I_M$ denotes an $M \times M$ unit matrix, and $\sigma_i(i=1,2,3)$ stands for the Pauli matrices. 
The boundary vector can be arranged as
\begin{eqnarray}
	\Phi_{\varphi}= \left(
		\begin{array}{c}
			\varphi(l_0-\epsilon) \\
			\varphi(l_1+\epsilon) \\
			(\CP_n \varphi) (l_0-\epsilon) \\
			(\CP_n \varphi) (l_1+\epsilon) \\
			(\CP_{n-1} \varphi) (l_0-\epsilon) \\
			(\CP_{n-1} \varphi) (l_1+\epsilon) \\
			(\CP_{n-1} \CP_n  \varphi) (l_0-\epsilon) \\
			(\CP_{n-1}\CP_n  \varphi) (l_1+\epsilon) \\
			(\CP_{n-2} \varphi) (l_0-\epsilon) \\
			(\CP_{n-2} \varphi) (l_1+\epsilon) \\
			(\CP_{n-2} \CP_n  \varphi) (l_0-\epsilon) \\
			(\CP_{n-2}\CP_n  \varphi) (l_1+\epsilon) \\
			(\CP_{n-2} \CP_{n-1}  \varphi) (l_0-\epsilon) \\
			(\CP_{n-2}\CP_{n-1}  \varphi) (l_1+\epsilon) \\
			(\CP_{n-2} \CP_{n-1} \CP_n  \varphi) (l_0-\epsilon) \\
			(\CP_{n-2}\CP_{n-1}  \varphi) (l_1+\epsilon) \\
			\vdots  \\
			(\CP_2 \CP_3 \cdots \CP_n \varphi) (l_0-\epsilon) \\
			(\CP_2 \CP_3 \cdots \CP_n \varphi) (l_1+\epsilon) \\
			\cdots \cdots \cdots \cdots \cdots \\
			\CP_1 \left[
				\begin{array}{c}
					{\rm upper \ half} \\
					{\rm components} 
				\end{array}
			\right]
		\end{array}
	\right).
\end{eqnarray}

For instance, 
$\Phi_{\varphi}$ for the case of $n=1$ (two singular points are placed at $x=0,l$)  is arranged as
\begin{eqnarray}
	\Phi_{\varphi} &=&\left(
		\begin{array}{c}
			\varphi(l-\epsilon) \\
			\varphi(0+\epsilon)\\
			\varphi(-l+ \epsilon)\\
			\varphi(0-\epsilon)
		\end{array}
	\right)
	=
	\left(
		\begin{array}{c}
			\varphi(l-\epsilon) \\
			\varphi(0+\epsilon) \\
			(\CP_1 \varphi)(l- \epsilon)\\
			(\CP_1 \varphi)(0+\epsilon)
		\end{array}
	\right),
\end{eqnarray}
which obeys the relations (\ref{B1})-(\ref{B3}) with $n=1$,
\begin{eqnarray}
	\Phi_{\CP_1 \varphi}&=&
		\left(
			\begin{array}{c}
				\left( \CP_1 \varphi\right)(l-\epsilon) \\
				\left( \CP_1 \varphi\right)(0+\epsilon) \\
				\left( \CP_1 \varphi\right)(-l+\epsilon) \\
				\left( \CP_1 \varphi\right)(0-\epsilon) 
			\end{array}
		\right)
		=
		\left(
			\begin{array}{c}
				 \varphi(-l+\epsilon) \\
				 \varphi(0-\epsilon) \\
				 \varphi(l-\epsilon) \\
				 \varphi(0+\epsilon) 
			\end{array}
		\right)
		=
		\left(
			\begin{array}{cc}
				0 & I_2 \\
				I_2 & 0
			\end{array}
		\right)
		\Phi_{\varphi}
		=\left( \sigma_1 \otimes I_2 \right)\Phi_{\varphi}, \nonumber  \\
	\\
	\Phi_{\CR_1 \varphi}&=&
		\left(
			\begin{array}{c}
				\left( \CR_1 \varphi\right)(l-\epsilon) \\
				\left( \CR_1 \varphi\right)(0+\epsilon) \\
				\left( \CR_1 \varphi\right)(-l+\epsilon) \\
				\left( \CR_1 \varphi\right)(0-\epsilon) 
			\end{array}
		\right)
		=
		\left(
			\begin{array}{c}
				 \varphi(l-\epsilon) \\
				 \varphi(0+\epsilon) \\
				- \varphi(-l+\epsilon) \\
				- \varphi(0-\epsilon) 
			\end{array}
		\right)
		=
		\left(
			\begin{array}{cc}
				I_2 & 0 \\
				0 & -I_2
			\end{array}
		\right)
		\Phi_{\varphi}
		=\left( \sigma_3 \otimes I_2 \right)\Phi_{\varphi}, \nonumber  \\
	\\
	\Phi_{\CQ_1 \varphi}&=&
		\left(
			\begin{array}{c}
				\left( \CQ_1 \varphi\right)(l-\epsilon) \\
				\left( \CQ_1 \varphi\right)(0+\epsilon) \\
				\left( \CQ_1 \varphi\right)(-l+\epsilon) \\
				\left( \CQ_1 \varphi\right)(0-\epsilon) 
			\end{array}
		\right)
		=
		\left(
			\begin{array}{c}
				 -i\varphi(-l+\epsilon) \\
				 -i\varphi(0-\epsilon) \\
				 i\varphi(l-\epsilon) \\
				 i\varphi(0+\epsilon) 
			\end{array}
		\right)
		=
		\left(
			\begin{array}{cc}
				0 & -i I_2 \\
				i I_2 & 0
			\end{array}
		\right)
		\Phi_{\varphi}
		=\left( \sigma_2 \otimes I_2 \right)\Phi_{\varphi}.  \nonumber \\*
\end{eqnarray}

Since our model has the singular points, wavefunction will, in general, be discontinuous there, but the discontinuity has to be controlled by the connection conditions that make the Hamiltonian hermitian. 
The hermiticity condition is 
\begin{eqnarray}
	\int_{-l}^l  dx \psi^* (x) \left( H^{A(B)} \varphi \right)(x) = \int_{-l}^l dx \left( H^{A(B)} \psi \right)^* (x) \varphi(x) \label{Hermitian}
\end{eqnarray}
for any wavefunctions $\psi(x), \varphi(x)$, 
where the integral $\int_{-l}^{l} dx$ is defined by
\begin{eqnarray}
	\int_{-l}^l dx\equiv \sum_{s=1}^{2^n}\int_{l_s+\epsilon}^{l_{s-1}-\epsilon} dx   \quad {\rm with}\  l_{2^n} \equiv l_0. 
\end{eqnarray}
In terms of the boundary vector, the requirement (\ref{Hermitian}) can simply be rewritten as the constraints on the boundary vector
\begin{equation}
	\Phi^{\dagger}_{\psi} \Phi_{\CD^{A(B)} \varphi}=\Phi^{\dagger}_{\CD^{A(B)}\psi} \Phi_{\varphi}. \label{H2}
\end{equation} 
In order to derive it, 
we have used the relation (\ref{B3}) and the formula of integration by parts
\begin{eqnarray}
	\int_{-l}^l dx \xi ^* (x) \left( \frac{d}{dx} \eta(x)\right) =-
		\int_{-l}^l dx \left( \frac{d}{dx} \xi (x) \right) ^* \eta(x) +\Phi_{\xi}^{\dagger} (\sigma_3 \otimes \cdots \otimes \sigma_3)\Phi_{\eta}, \label{integ_by_parts}
\end{eqnarray}
where the functions $\xi(x)$ and $\eta(x)$ are assumed to be continuous everywhere except for the singular points.
It is easy to show that eq. (\ref{H2}) is equivalent to 
\begin{eqnarray}
	| \Phi _{\varphi} -iL_0 \Phi_{\CD^{A(B)} \varphi} | = |\Phi_{\varphi} +i L_0 \Phi_{\CD^{A(B)} \varphi} |,  \label{H3}
\end{eqnarray}
where $L_0$ is an arbitrary nonzero constant with the dimension of length. 
Then, the condition (\ref{H3}) can be written as 
\begin{equation}
	\left(
		I_{2^{n+1}}- U 
	\right)\Phi_{\varphi} +i L_0 
	\left(
		I_{2^{n+1}}+U
	\right)\Phi_{\CD^{A(B)} \varphi} =0,
	\label{CC}
\end{equation}
where $U$ is a $2^{n+1} \times 2^{n+1}$ unitary matrix. 
Thus, we have found that the connection conditions which make the Hamiltonian hermitian are given by eq. (\ref{CC}). 
The $2^n$ singularities in our model are characterized by a $2^{n+1}\times 2^{n+1}$ unitary matrix, $U$.

It is important to notice that the above connection condition does not necessarily guarantee the $N=2n+1$ supersymmetry, because the hermiticity of the Hamiltonian does not, in general, ensure that of the 
supercharges. 
Moreover, the state, $\left(Q_a \varphi\right)(x)$ does not necessarily  belong to the same functional space as $\varphi(x)$.
This also means that $\left(Q_a \varphi\right)(x)$ does not obey the same connection conditions as $\varphi(x)$.
In order for any operator ${\cal O}$ to be physical, the state $\left( {\cal O}\varphi\right)(x)$ for any state $\varphi(x)$ has to obey the same connection conditions as $\varphi(x)$.
In the following, we call ${\cal O}$ a physical operator if ${\cal O}$ is hermitian {\it and} $\left( {\cal O}\varphi \right)(x)$ satisfies the same connection conditions as $\varphi(x)$.

In the standard argument of supersymmetry,
for any energy eigenfunction $\varphi$ with an energy $E$, $Q_a \varphi$ is also an energy eigenfunction with the same energy $E(>0)$
and then $Q_a \varphi$ is called a supersymmetric partner of $\varphi$. 
In our model, this is true  
if and only if $Q_a$ is a physical operator, otherwise the state $Q_a \varphi$ should be removed from the functional space, 
in which any state satisfies the connection conditions (\ref{CC}) with a characteristic matrix $U$. 
Therefore the $N=2n+1$ supersymmetry can be realized in our model, only when all the $N=2n+1$ supercharges are physical. 
We will see below that this requirement severely restricts a class of the connection conditions. 

Let $\varphi(x)$ be any wavefunction obeying 
the connection conditions (\ref{CC}), and satisfying Schr{\" o}dinger equation, $H^{A(B)} \varphi(x) =E \varphi(x)$.

First, we require that $(Q_a^{A(B)} \varphi )(x)$ $(a=1,2,\cdots, 2n+1)$ obey the same connection conditions as $\varphi(x)$, 
\begin{equation}
	\left(
		I_{2^{n+1}}- U 
	\right)\Phi_{Q_{a}^{A(B)} \varphi} +i L_0 
	\left(
		I_{2^{n+1}}+U
	\right)\Phi_{\CD^{A(B)} (Q_{a}^{A(B)} \varphi)} =0. \label{H4}
\end{equation}
Here, $\varphi$ in eq. (\ref{CC}) has been replaced by $Q_{a}^{A(B)}\varphi$.
By noting (\ref{Q_A}), (\ref{Q_B}), (\ref{GD}), (\ref{SuperA1})-(\ref{SuperA2}) and 
$H^{A(B)} \varphi(x) =E \varphi(x)$, eq. (\ref{H4}) leads to
\begin{equation}
	\left(
		I_{2^{n+1}}- U 
	\right)\Phi_{\Gamma_a \CD^{A(B)} \varphi} - 2i L_0 
	E \left(
		I_{2^{n+1}}+U
	\right)\Phi_{\Gamma_a \varphi} =0. \label{H5}
\end{equation} 
The connection conditions must be independent of the energy $E$, so that we find the eigenvalues of $U$ must be $\pm1$, 
i.e. $U^2=I_{2^{n+1}}$.
Then, (\ref{H5}) can be written as 
\begin{eqnarray}
&& (I_{2^{n+1}} -U ) \Phi_{\Gamma_a \CD^{A(B)} \varphi} =0, \label{Q1}\\
&& (I_{2^{n+1}} +U ) \Phi_{\Gamma_a \varphi}=0. \label{Q2}
\end{eqnarray}
We note that the case of $U =\pm I_{2^{n+1}}$ turns out to lead to no nontrivial models because all wavefunctions would vanish.
It is convenient to introduce $\gamma_a$ $(a=1,2,\cdots, 2n+1)$ as
\footnote{The definition of $\gamma_a$ for $a=1,2,\cdots, 2n$ is slightly different from those in ref. \cite{susy4}.}
\begin{eqnarray}
	\gamma_{2k-1} &=&\left( \vec{e}_{\CR_1} \cdot \vec{\sigma} \otimes \cdots \otimes \vec{e}_{\CR_{k-1}} \cdot \vec{\sigma} \otimes \vec{e}_{\CP_k} \cdot \vec{\sigma} \otimes I_2 \otimes \cdots \otimes I_2 \otimes \sigma_3 \right) ,\\
	\gamma_{2k} &=&\left( \vec{e}_{\CR_1} \cdot \vec{\sigma} \otimes \cdots \otimes \vec{e}_{\CR_{k-1}} \cdot \vec{\sigma} \otimes \vec{e}_{\CQ_k} \cdot \vec{\sigma} \otimes I_2 \otimes \cdots \otimes I_2 \otimes \sigma_3 \right), \\
	\gamma_{2n+1} &=& \left( I_2 \otimes \cdots \otimes I_2 \otimes \sigma_2\right), 
	\qquad \qquad  \qquad \qquad \quad 
	k=1,2,\cdots, n,
\end{eqnarray}
where $\vec{\sigma} = (\sigma_1 ,\sigma_2 ,\sigma_3)$.  
The $\gamma_{a}$ for $a=1,2,\cdots,2n+1$ satisfy the same anticommutation relations as $\Gamma_{a}$, 
\begin{eqnarray}
	\{\gamma_a, \gamma_b \}= 2 \delta_{ab}.
\end{eqnarray}
Using these, we have the relation
\begin{equation}  
\Phi_{\Gamma_a \varphi} = \gamma_a \Phi_{\varphi},
\end{equation}
and then, 
eqs. (\ref{Q1}), (\ref{Q2}) become
\begin{eqnarray}
	&&(I_{2^{n+1}}-\gamma_a U \gamma_a ) \Phi_{\CD^{A(B)}\varphi}= 0,\label{Q3}\\
	&&(I_{2^{n+1}}+\gamma_a U \gamma_a) \Phi_{\varphi}=0. \label{Q4}
\end{eqnarray}
Eq. (\ref{CC}) is reduced to 
\begin{eqnarray}
	&&(I_{2^{n+1}}-U) \Phi_{\varphi}= 0,\\
	&&(I_{2^{n+1}}+U)  \Phi_{\CD^{A(B)}\varphi}=0,
\end{eqnarray}
under the condition of $U^2=I_{2^{n+1}}$.
Since our model contains the $2^n$ point singularities, we need to impose $2\times 2^n$ connection conditions at the singularities to solve the Schr{\" o}dinger equation.
If the number of conditions is larger than $2\times 2^n$, the model 
would become trivial due to overconstraints.
Since the number of the conditions (\ref{CC}), which make the Hamiltonian hermitian, is $2\times 2^n$,  eqs. (\ref{Q3}) and (\ref{Q4}) should produce no new constraints. 
This implies that the characteristic matrix $U$ has to satisfy 
\begin{equation}
	\gamma_a U \gamma_a =- U,  \qquad  a=1,2,\cdots, 2n+1.
\end{equation}
Thus, we have found the constraints on the characteristic matrix $U$,
\begin{eqnarray}
	U^{\dagger}U &=& I_{2^{n+1}}, \\
	U^2 &=&I_{2^{n+1}}, \\
	\gamma_a U \gamma_a &=&- U, \qquad 
	a=1,2,\cdots,2n+1 \label{C1}
\end{eqnarray}
which guarantee that the Hamiltonian is hermitian and $(Q_a^{A(B)}\varphi)(x)$ $(a=1,2,\cdots,2n+1)$ satisfy the same connection conditions as $\varphi(x)$.

Before we discuss the hermiticity of the supercharges, let us determine the form of the characteristic matrix $U$.
An arbitrary $2^{n+1} \times 2^{n+1}$ matrix can be uniquely expanded as
\begin{eqnarray}
	\sum_{\alpha_1 =0}^3 \cdots \sum_{\alpha_{n+1}=0}^3 C_{\alpha_1,\alpha_2, \cdots, \alpha_{n+1}}\left(  \tilde{e}_{\alpha_1}^{(1)}\otimes \tilde{e}_{\alpha_2}^{(2)}\otimes \cdots \otimes \tilde{e}_{\alpha_n}^{(n)}\otimes \tilde{e}_{\alpha_{n+1}}^{(n+1)}
	\right), \label{A_U1}
\end{eqnarray}
where
\begin{eqnarray}
	&&\tilde{e}_{\alpha_k=0}^{(k)} \equiv I_2,\ 
	\tilde{e}_{\alpha_k=1}^{(k)} \equiv \vec{e}_{\CP_k}\cdot \vec{\sigma},\  
	 \tilde{e}_{\alpha_k=2}^{(k)} \equiv \vec{e}_{\CQ_k}\cdot \vec{\sigma}, \ 
	\tilde{e}_{\alpha_k=3}^{(k)} \equiv \vec{e}_{\CR_k}\cdot \vec{\sigma}, \nonumber \\
	&& \qquad \qquad \qquad \qquad \qquad \qquad \qquad  \qquad \qquad 
	 k=1,2,\cdots, n, \label{A_U2}
\end{eqnarray}
and 
\begin{eqnarray}
	&&\tilde{e}_{\alpha_{n+1}=0}^{(n+1)} \equiv I_2,  \ 
	\tilde{e}_{\alpha_{n+1}=1}^{(n+1)} \equiv \sigma_1,\  
	\tilde{e}_{\alpha_{n+1}=2}^{(n+1)} \equiv \sigma_2,  \ 
	  \tilde{e}_{\alpha_{n+1}=3}^{(n+1)} \equiv \sigma_3.  \label{A_U3}
\end{eqnarray}
First, we impose the conditions (\ref{C1}) on the characteristic matrix $U$.
Then, some coefficients in the expansion 
vanish except for $C_{\alpha_1=0, \cdots, \alpha_n=0, \alpha_{n+1}= 1}$, $C_{\alpha_1=3, \cdots, \alpha_n=3, \alpha_{n+1}=3}$.
Thus, the characteristic matrix $U$ becomes  
\begin{eqnarray}
	U&=& b_1
	 \left(
		I_2 \otimes \cdots I_2 \otimes \sigma_1 
	\right) 
	+ b_2
	\left(
		\vec{e}_{\CR_1} \cdot \vec{\sigma} \otimes \cdots \otimes \vec{e}_{\CR_n}\cdot \vec{\sigma} \otimes \sigma_3
	\right),
\end{eqnarray}
where 
\begin{eqnarray}
	b_1\equiv C_{\alpha_1=0, \cdots, \alpha_n=0, \alpha_{n+1}= 1}, \quad 
	b_2\equiv C_{\alpha_1=3, \cdots, \alpha_n=3, \alpha_{n+1}=3}.
\end{eqnarray}
The conditions $U^{\dagger}U=I_{2^{n+1}}$ and $U^2=I_{2^{n+1}}$ imply $ U^{\dagger}=U$, so that we have 
\begin{eqnarray}
	b_1, b_2 \in {\bf R}, \quad 	\left( b_1 \right) ^2 + \left( b_2 \right)^2=1.
\end{eqnarray}
Therefore, the connection conditions can be written by
\begin{eqnarray}
	&&\left( I_{2^{n+1}} -U(b_1,b_2) \right) \Phi_{\varphi} =0, \label{CC1}\\
	&&\left( I_{2^{n+1}}+U(b_1,b_2) \right) \Phi_{\CD^{A(B)} \varphi}=0, \label{CC2}
\end{eqnarray}
where 
\begin{eqnarray}
		U(b_1,b_2)=b_1 \left(
		I_2 \otimes \cdots I_2 \otimes \sigma_1 
	\right)
	+b_2
	\left(
		\vec{e}_{\CR_1} \cdot \vec{\sigma} \otimes \cdots \otimes \vec{e}_{\CR_n}\cdot \vec{\sigma} \otimes \sigma_3
	\right)
\end{eqnarray}
with $b_1, b_2 \in {\bf R}$ and $(b_1)^2+(b_2)^2=1$.

Now, an important question that remains to be answered is whether the supercharges are hermitian under the connection conditions (\ref{CC1}) and (\ref{CC2}) or not.
Using (\ref{integ_by_parts}), we have the relation 
\begin{eqnarray}
	&&\int_{-l}^l dx \psi^*(x) \left( Q_a \varphi \right)(x) =\int_{-l}^l dx \left( Q_a \psi \right) ^* (x) \varphi(x)+ \frac{i}{2} \Phi_{\psi}^{\dagger}\gamma_a \Phi_{\varphi}, \nonumber \\
	&&\qquad \qquad \qquad \qquad \qquad \qquad \qquad \qquad \qquad a=1,2,\cdots,2n+1. \label{hermitian_Q}
\end{eqnarray}
It is not difficult to show that the surface term in (\ref{hermitian_Q}) vanishes, 
\begin{eqnarray}
	\Phi_{\psi}^{\dagger} \gamma_a \Phi_{\varphi}=0, \qquad  a=1,2,\cdots, 2n+1, \label{hermite}
\end{eqnarray}
if wavefunctions $\varphi$ and $\psi$ obey the connection conditions (\ref{CC1}) and (\ref{CC2}). 
Thus, the connection conditions (\ref{CC1}) and (\ref{CC2}) also make all the supercharges hermitian.

Thus, we have found that the connection conditions, which make not only the Hamiltonian hermitian but also the $2n+1$ supercharges physical, are given by eq. (\ref{CC1}), (\ref{CC2}).
Then, the $N=2n+1$ supersymmetry can be realized with  
the connection conditions (\ref{CC1}), (\ref{CC2}).
\subsection{Degeneracy of the spectrum}

In this section, we study the degeneracy of the spectrum in the $N=2n+1$ supersymmetric models, in particular, vacuum states with the vanishing energy.
We consider the two types of the $N=2n+1$ supersymmetric models  (type A and type B), separately.
\\
\\
{\bf 3.3.1  Type A model}
\\\\
We first consider the degeneracy in the type A supersymmetric model.
The connection conditions compatible with the $N=2n+1$ supercharges $Q_{1}^{A}, \cdots, Q_{2n+1}^{A}$ are given by
\begin{eqnarray}
	&&(I_{2^{n+1}}-U(b_1,b_2)) \Phi_{\varphi}=0, \label{A_CC1}\\
	&& (I_{2^{n+1}}+U(b_1,b_2)) \Phi_{\CD^A \varphi}=0 , \label{A_CC2}
\end{eqnarray}
where
\begin{eqnarray}
	U(b_1,b_2)=b_1 (I_2 \otimes \cdots \otimes I_2 \otimes \sigma_1 ) +b_2 (\vec{e}_{\CR_1} \cdot \vec{\sigma} \otimes \cdots \otimes \vec{e}_{\CR_n} \cdot \vec{\sigma}\otimes \sigma_3 )
\end{eqnarray}
with $b_1, b_2 \in {\bf R}$ and $(b_1)^2+(b_2)^2=1$.
We note that the $\CG_{\CR_k} \ (k=1,2,\cdots, n)$ commutes with the Hamiltonian $H^A$ and also with each other.
Since they are also physical under the connection conditions (\ref{A_CC1}) and (\ref{A_CC2}), we can introduce simultaneous eigenfunctions of $H^A$ and $\CG_{\CR_k} $ 
such that
\begin{eqnarray}
	&&H^A \varphi_{E;\lambda_1, \cdots, \lambda_n} (x) =E \varphi_{E;\lambda_1, \cdots, \lambda_n} (x), \\
	&& \CG_{\CR_k}\varphi_{E;\lambda_1, \cdots, \lambda_n} (x)=\lambda_k \varphi_{E;\lambda_1, \cdots, \lambda_n} (x) 
\end{eqnarray}
with $\lambda_k=1$ or $-1$ for $k=1,2, \cdots ,n$.
Since $Q_a^A \ (a=1,2,\cdots, 2n+1)$ and $\CG_{\CR_k}\ (k=1,2,\cdots, n)$ satisfy the relations
\begin{equation}
	Q_a^A \CG_{\CR_k} =\left\{
		\begin{array}{l}
			-\CG_{\CR_k} Q_a^A \quad {\rm for} \ a=2k-1\  {\rm or}\  2k, \\
			+\CG_{\CR_k} Q_a^A \quad {\rm otherwise},
		\end{array}
	\right.
\end{equation}
the states $Q_{2k-1}^{A}\varphi_{E;\lambda_1, \cdots, \lambda_n}(x)$ and $Q_{2k}^{A}\varphi_{E;\lambda_1, \cdots, \lambda_n}(x)$ should be proportional to \sloppy $\varphi_{E;\lambda_1, \cdots, -\lambda_k, \cdots, \lambda_n}(x)$, and the states $Q_{2n+1}^{A}\varphi_{E;\lambda_1, \cdots, \lambda_n}(x)$ should be proportional to $\varphi_{E;\lambda_1, \cdots, \lambda_n}(x)$. 
Noting that $Q_{2k-1}^{A}=-i Q_{2k}^A \CG_{\CR_k}$, we have the relations
\begin{eqnarray}
	&&Q_{2k-1}^A \varphi_{E;\lambda_1,\cdots, \lambda_n}(x) = -i \lambda_k Q_{2k}^A \varphi_{E;\lambda_1,\cdots, \lambda_n}(x) \propto \varphi_{E;\lambda_1,\cdots,-\lambda_k,\cdots, \lambda_n}(x), \label{A_2k-1_2k}\\
	&& Q_{2n+1}^A \varphi_{E;\lambda_1,\cdots, \lambda_n}(x) \propto \varphi_{E;\lambda_1,\cdots, \lambda_n}(x),
\end{eqnarray}
 when $E > 0$.
These imply that the degeneracy of the spectrum for $E> 0$ is given by $2^n$.
Some examples of the transition for the eigenfunction $\varphi_{E_1;\lambda_1, \cdots, \lambda_n}$ by acting the supercharges are depicted in Fig. 5.
\begin{figure}[h]
\begin{center}
	\includegraphics{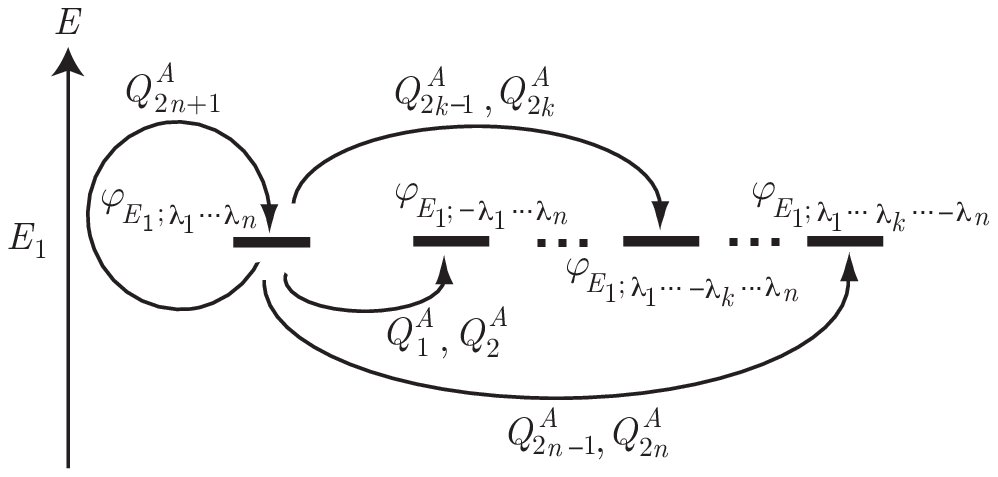}
\end{center}
\caption{Examples of the transition for the eigenfunction $\varphi_{E_1; \lambda_1, \cdots, \lambda_n} $ under the action of the supercharges ($Q_1^A, \cdots, Q_{2n+1}^A$)}
\end{figure}
This degeneracy  
can also be understood 
from an algebraic point of view;
for fixed nonzero energy $E$, $Q_a^A/\sqrt{E}$ for $a=1,2,\cdots,2n+1$ form the Clifford algebra in $(2n+1)$-dimensions, and the representation is known as $2^n$.

The above argument cannot apply for the case of $E=0$. 
 This is because any state $\varphi_0(x)$ with vanishing energy must satisfy $Q_a^A \varphi_0 (x)=0$ for $a=1,2, \cdots,2n+1$.
 The equations $Q_a^A \varphi_0 (x)=0$ are equivalent to
 \begin{equation}
 	\CD^A  \varphi_0 (x)=0 . \label{3.69}
\end{equation}
The above equation is easily solved, and we have  formal solutions as
\begin{eqnarray}
\varphi_{0;\lambda_1,\cdots, \lambda_n} (x) 
	&=& N_{\lambda_1, \cdots, \lambda_n}
		\left[
			\prod_{k=1}^{n}
				\frac{1}{2}\left(1+\lambda_k \CG_{\CR_k}\right)
		\right]
		e^{-\lambda_1\cdots \lambda_n W(x)}. \label{Z1}
\end{eqnarray}
Here, $N_{\lambda_1, \cdots, \lambda_n}$ 
denotes normalization constants.
By noting the Hamiltonian $H^A$ includes only $W'(x)$ (but not $W(x)$ itself), we have an ambiguity to choose $W(x)$ to satisfy
\begin{eqnarray}	
\CP_{k} W(x) \CP_k &=&W(x), \qquad  k=1,2,\cdots, n, \\
\CP_{n+1}W(x) \CP_{n+1} &=& -W(x).
\end{eqnarray}
For a noncompact space, any non-normalizable states would be removed from the functional space.
The space is, however, compact in our model, so that these solutions are always normalizable.
Nevertheless, some of them must be removed from the functional space.
This is because some solutions are incompatible with the connection conditions.
Although the zero energy states trivially satisfy the half of the connection conditions (\ref{A_CC2}), we need to verify whether the states (\ref{Z1}) satisfy the remaining connection condition (\ref{A_CC1}) or not.

In order to investigate the consistency of the zero energy states with the connection conditions, we use the transformation (\ref{singular}), (\ref{SINGULAR2}).
This transformation yields 
\begin{eqnarray}
	\tilde{U}(b_1,b_2) \equiv V U(b_1,b_2) V^{\dagger} =b_1\left( I_2 \otimes \cdots \otimes  I_2 \otimes \sigma_1 \right)+b_2 \left(\sigma_3 \otimes \cdots \otimes \sigma_3 \right),  \label{Singular}
\end{eqnarray}
where
\begin{eqnarray}
	V\equiv \left( e^{i\vec{v}_1 \cdot \vec{\sigma}} \otimes e^{i\vec{v}_2 \cdot \vec{\sigma}} \otimes \cdots \otimes e^{i\vec{v}_n\cdot \vec{\sigma}} \otimes I_2 \right), \qquad 
	e^{i\vec{v}_k\cdot\vec{\sigma}}\in SU(2). \label{Singular2}
\end{eqnarray}
The transformation is a {\it singular} unitary transformation because it, in general, changes connection conditions of wavefunctions at singular points.
The transformation may be regarded as a duality connecting different theories (with different connection conditions). Moreover, the transformation gives us the convenient expression for the connection condition. 

For instance, we consider the model which possesses two point singularities placed at $x=0, l$ ($n=1$) as an example.
The connection condition (\ref{A_CC1}) is transformed by (\ref{Singular2}) to
\begin{eqnarray}
	(I_4 - \tilde{U}(b_1,b_2) ) \Phi_{
	\tilde{\varphi}}=0, \label{A_CC3}
\end{eqnarray}
where 
\begin{eqnarray}
	\tilde{U}(b_1, b_2)= b_1 (I_2 \otimes \sigma_1) +b_2 (\sigma_3 \otimes \sigma_3), \quad 
	 \tilde{\varphi} =\CV_1 \varphi, \quad \CV_1=e^{i \vec{v}_1\cdot \vec{\CP}_1}.
\end{eqnarray}
Explicitly, eq. (\ref{A_CC3}) is written as 
\begin{eqnarray}
	\left( 
	\begin{array}{cccc}
		1-b_2 & -b_1 &0&0 \\
		-b_1 &1+b_2 &0&0 \\
		0&0& 1+b_2& -b_1 \\
		0&0& -b_1 &1-b_2 
	\end{array}
	\right)
	\left( 
		\begin{array}{c}
			\tilde{\varphi}(l-\epsilon)\\
			\tilde{\varphi}(0+\epsilon) \\
			 \tilde{\varphi}(-l+\epsilon) \\
			\tilde{\varphi}(-0-\epsilon)
		\end{array}
	\right)
	=0.
\end{eqnarray}
We observe that the above connection condition is separated into two
independent conditions, depending on the argument of the wavefunction. This suggests that it is useful to assign the eigenvalue $\lambda_1$ of the operator $\CR_1$ for the wavefunction. To this end, we define  the projection operator,
\begin{equation}
{1\over 2}\left( 1+\lambda_1\CR_1 \right).
\end{equation}
By multiplying $(1+\lambda_1\CR_1)/2$ from the left to eq. (\ref{A_CC3}), we obtain the boundary vector as
\begin{eqnarray}
	\Phi_{\left(\frac{1+\lambda_1 \CR_1}{2}\right) \tilde{\varphi}}
	&=&
	\left\{ 
	\begin{array}{ll}
	\left( 
		\begin{array}{c}
			\tilde{\varphi}_{\lambda_1=+1}(l-\epsilon)=0 \\
			\tilde{\varphi}_{\lambda_1=+1}(0+\epsilon)=0 \\
			0 \\
			0
		\end{array}
	\right)& {\rm for}\ \lambda_1=+1, \\ 	
	\left(
				\begin{array}{c}
				0\\
				0\\
					\tilde{\varphi}_{\lambda_1=-1}(-l+\epsilon)=0 \\
					\tilde{\varphi}_{\lambda_1=-1}(0-\epsilon)=0 
				\end{array}
			\right)& {\rm for}\ \lambda_1=-1. 
	\end{array}
	\right.
\end{eqnarray}
Then, the connection conditions (\ref{A_CC3}) are reduced to 
\begin{itemize}
	\item for $b_2=1$ (and $b_1=0$)
\end{itemize}
\begin{eqnarray}
	\tilde{\varphi}_{\lambda_1=+1}(0+\epsilon)=0, \label{A_CC4} \\
	\tilde{\varphi}_{\lambda_1=-1}(-l+\epsilon)=0, \label{A_CC5}
\end{eqnarray}
\begin{itemize}
	\item for $b_2\ne1$
\end{itemize}
\begin{eqnarray}
	(1-b_2) \tilde{\varphi}_{\lambda_1=+1}(l-\epsilon)-b_1 \tilde{\varphi}_{\lambda_1=+1}(0+\epsilon)=0, \label{A_CC6}\\ 
	(1-b_2) \tilde{\varphi}_{\lambda_1=-1}(0-\epsilon)-b_1 \tilde{\varphi}_{\lambda_1=-1}(-l+\epsilon)=0. \label{A_CC7}
\end{eqnarray}
Let us also note that the Schr{\" o}dinger equations can be written separately in each region of $(0<x<l)$ and $(-l<x<0)$, 
\begin{eqnarray}
	&&\tilde{H} \tilde{\varphi}_{\lambda_1=+1}(x) = E \tilde{\varphi}_{\lambda_1=+1}(x) \quad (0 <x<l) \qquad {\rm with}\ (\ref{A_CC4}) \ {\rm or}\ (\ref{A_CC6}), \\
	&&\tilde{H} \tilde{\varphi}_{\lambda_1=-1}(x) = E \tilde{\varphi}_{\lambda_1=-1}(x) \quad (-l<x<0)\qquad {\rm with}\ (\ref{A_CC5}) \ {\rm or}\ (\ref{A_CC7}), 
\end{eqnarray}
where $\tilde{H} \equiv \CV H \CV^{\dagger}$. 

Similarly, we can generalize the above argument to the case of $n$. The connection conditions (\ref{A_CC1}) 
can be written as
\begin{eqnarray}
	(I_{2^{n+1}}-\tilde{U}(b_1,b_2)) \Phi_{ \left(\prod_{k=1}^{n} \frac{1+\lambda_k \CR_k}{2} \right)\tilde{\varphi}}=0 \label{A_CC8}
\end{eqnarray}
where
\begin{eqnarray}
	\tilde{\varphi} \equiv \CV \varphi \quad 
{\rm with}\quad  \CV \equiv \CV_1 \CV_2 \cdots \CV_n, \ \CV_i=e^{i \vec{v}_i \cdot \vec{\CP_i}}. 
 \label{Singular3}
\end{eqnarray} 
In this basis, 
the boundary vector can simply be written as
\begin{eqnarray}
	\Phi_{\left( \prod_{k=1}^{n} \frac{1+\lambda_k \CR_k}{2}  \right)\tilde{\varphi}} 
		=
	\Phi_{\tilde{\varphi}_{\lambda_1,\cdots,\lambda_n}}
	&=&
	\left(
		\begin{array}{c}
			0 \\
			\vdots \\
			0\\
			\left( \CP_1^{\frac{1-\lambda_1}{2}} \cdots\CP_n^{\frac{1-\lambda_n}{2}}\tilde{\varphi}_{\lambda_1,\cdots,\lambda_n}\right)(l_0-\epsilon) \\
			\left( \CP_1^{\frac{1-\lambda_1}{2}} \cdots\CP_n^{\frac{1-\lambda_n}{2}}\tilde{\varphi}_{\lambda_1,\cdots,\lambda_n}\right)(l_1+\epsilon) \\
			0\\
			\vdots \\
			0
		\end{array}
	\right). 
\end{eqnarray}
The connection conditions (\ref{A_CC8}) are reduced to 
\begin{itemize}
	\item {\rm for \ $b_2=1$ (and $b_1=0$)} 
\end{itemize}
\begin{eqnarray}
		&&\left( \CP_1^{\frac{1-\lambda_1}{2}} \cdots\CP_n^{\frac{1-\lambda_n}{2}}\tilde{\varphi}_{\lambda_1,\cdots,\lambda_n}\right)(l_0-\epsilon)=0 
 \qquad {\rm for} \ \lambda_1\cdots \lambda_n =-1, \label{A_cc3}\\
		&&\left( \CP_1^{\frac{1-\lambda_1}{2}} \cdots\CP_n^{\frac{1-\lambda_n}{2}}\tilde{\varphi}_{\lambda_1,\cdots,\lambda_n}\right)(l_1+\epsilon)=0 
		 \qquad {\rm for} \ \lambda_1\cdots \lambda_n =+1, \label{A_cc4}
\end{eqnarray}
\begin{itemize}
	\item {\rm  for $b_2\ne 1$}
\end{itemize}
\begin{eqnarray}
	&&	(1-b_2) \left( \CP_1^{\frac{1-\lambda_1}{2}} \cdots\CP_n^{\frac{1-\lambda_n}{2}}\tilde{\varphi}_{\lambda_1,\cdots,\lambda_n}\right)(l_0-\epsilon) -b_1\left( \CP_1^{\frac{1-\lambda_1}{2}} \cdots\CP_n^{\frac{1-\lambda_n}{2}}\tilde{\varphi}_{\lambda_1,\cdots,\lambda_n}\right)(l_1+\epsilon)=0 \nonumber \\ &&
\qquad \qquad \qquad \qquad \qquad  \qquad \qquad \qquad \qquad \qquad \qquad {\rm for}\ \lambda_1\cdots\lambda_n=+1, \label{A_cc1} \\
	&&	(1-b_2)\left( \CP_1^{\frac{1-\lambda_1}{2}} \cdots\CP_n^{\frac{1-\lambda_n}{2}}\tilde{\varphi}_{\lambda_1,\cdots,\lambda_n}\right)(l_1+\epsilon) -b_1\left( \CP_1^{\frac{1-\lambda_1}{2}} \cdots\CP_n^{\frac{1-\lambda_n}{2}}\tilde{\varphi}_{\lambda_1,\cdots,\lambda_n}\right)(l_0-\epsilon)=0  \nonumber \\
	&& \qquad \qquad \qquad \qquad \qquad \qquad \qquad \qquad \qquad \qquad \qquad {\rm for}\ \lambda_1\cdots\lambda_n=-1. \label{A_cc2}
\end{eqnarray}
The Schr{\" o}dinger equations can be separated into $2^n$ regions of $(l_s <x < l_{s+1})$ $(s=0,1,\cdots,2^n-1)$.

Now, we are ready to discuss  whether the zero energy states
satisfy the connection conditions or not.
With the transformations (\ref{Singular3}), the zero energy states (\ref{Z1}) become 
\begin{eqnarray}
	\tilde{\varphi}_{0;\lambda_1,\cdots, \lambda_n}(x)\equiv \CV \varphi_{0;\lambda_1,\cdots, \lambda_n}(x) =N_{\lambda_1, \cdots, \lambda_n}\left[
		\prod_{k=1}^{n} \frac{1}{2} ( 1+\lambda_k\CR_k) \right]e^{-\lambda_1\cdots \lambda_n W(x)}. \label{Z2}
\end{eqnarray}
The states (\ref{Z2}) do not satisfy the connection conditions (\ref{A_cc3}), (\ref{A_cc4}) for $b_2=1$.
Although for general $W(x)$, the connection conditions (\ref{A_cc1}), (\ref{A_cc2}) are not satisfied by the zero energy states for $b_2 \ne 1$, there
is, however, an exceptional case, where $W(x)$ satisfies
\begin{eqnarray}
	\sqrt{\frac{1-b_1}{1+b_1}}=e^{2 W(l_0-\epsilon)}. \label{S1}
\end{eqnarray}
With this special $W(x)$, all the states (\ref{Z2}) become supersymmetric vacuum states compatible with the connection conditions.

There is no supersymmetric vacuum state for general $W(x)$.
Therefore, the supersymmetry is \lq\lq spontaneously"\footnote{
If there is no zero energy state, we say that supersymmetry is spontaneously broken by analogy with supersymmetric quantum field theory.
Other mechanisms of (spontaneous) supersymmetry breaking due to boundary effects have been found in Refs. \cite{STT,Takenaga}.
}
broken.
While, for the special $W(x)$ satisfying (\ref{S1}), we have the supersymmetric vacuum states, and the degeneracy is $2^{n}$.

Let us note that the $\CP_{n+1}$ becomes physical when $b_2=0$
 and commutes with the Hamiltonian $H^A$ only for $W''(x)=0$.
 We can introduce the simultaneous eigenfunctions of the Hamiltonian $H^A$, $\CG_{\CR_k}$ $(k=1,2,\cdots,n)$ and $\CP_{n+1}$,
\begin{eqnarray}
	H^A \varphi_{E;\lambda_1,\cdots, \lambda_n, \lambda_{n+1}}(x)&=&E\varphi_{E;\lambda_1,\cdots, \lambda_n, \lambda_{n+1}}(x), \\
	\CG_{\CR_k}\varphi_{E;\lambda_1,\cdots, \lambda_n, \lambda_{n+1}}(x) &=& \lambda_k \varphi_{E;\lambda_1,\cdots, \lambda_n, \lambda_{n+1}}(x), \\
	\CP_{n+1} \varphi_{E;\lambda_1,\cdots, \lambda_n, \lambda_{n+1}}(x)&=& \lambda_{n+1}\varphi_{E;\lambda_1,\cdots, \lambda_n, \lambda_{n+1}}(x). 
\end{eqnarray}

The new label $\lambda_{n+1}$ on the wavefunction suggests the enhancement of the degeneracy for the state with $E>0$. 
In order to confirm it, 
let us define new operator $\hat{Q}_{a}^{A}\ (a=1,2,\cdots,2n+1)$ as
\begin{equation}
	\hat{Q}_{a}^{A} \equiv \frac{i}{2}\Gamma_a \left(\CR_1 \cdots \CR_{n+1} \frac{d}{dx}\right).
\end{equation}
Then, the Hamiltonian $H^A$ can be written as
\begin{eqnarray}
	H^A &=&2(Q_a^A)^2= 2(\hat{Q}_{a}^{A})^2 +\frac{1}{2} c^2, 
\end{eqnarray}
where $c^2\equiv (W'(x))^2$ is independent of $x$ for $W''(x)=0$. The energy is bounded from below as 
\begin{equation}
	E  \ge \frac{1}{2} c^2. 
\end{equation}

We can show that ${\hat{Q}_a^A}$ $(a=1,2,\cdots, 2n+1)$ is physical under the connection conditions (\ref{A_CC1}) and (\ref{A_CC2}).
Since $\hat{Q}_{a}^{A}$ and $\CG_{\CR_k}$ ($k=1,2,\cdots,n$), $\CP_{n+1}$ satisfy 
\begin{eqnarray}
	&&\CP_{n+1} \hat{Q}_{a}^{A} =- \hat{Q}_{a}^{A} \CP_{n+1} \quad \quad {\rm for} \ a=1,2,\cdots, 2n+1.\\
	&&\CG_{\CR_k} \hat{Q}_{a}^{A}=
		\left\{ \begin{array}{l}
			-\hat{Q}_{a}^{A}\CG_{\CR_k}\qquad {\rm for}\ a=2k-1,2k,\\
			+\hat{Q}_{a}^{A}\CG_{\CR_k} \qquad {\rm otherwise}
		\end{array}
		\right. 
\end{eqnarray}
and $\hat{Q}_{2k-1}^A =-i \hat{Q}_{2k}^A \CG_{\CR_k}$,
we have the relations 
\begin{eqnarray}
	&&\hat{Q}_{2k-1}^{A} \varphi_{E;\lambda_1,\cdots, \lambda_n, \lambda_{n+1}}=-i\lambda_k \hat{Q}_{2k}^{A} \varphi_{E;\lambda_1,\cdots, \lambda_n, \lambda_{n+1}} \propto \varphi_{E;\lambda_1,\cdots,-\lambda_k, \cdots, \lambda_n, -\lambda_{n+1}}, \\
	&&\hat{Q}_{2n+1}^{A} \varphi_{E;\lambda_1,\cdots, \lambda_n, \lambda_{n+1}} \propto \varphi_{E;\lambda_1,\cdots,\lambda_n, -\lambda_{n+1}}.
\end{eqnarray}
Thus, in this case, the degeneracy for $E> \frac{1}{2} c^2$ is given by $2^{n+1}$.
This result can also be 
obtained from an algebraic point of view;
 $\hat{Q}_{a}^{A} / \sqrt{E}$ $(a=1,2,\cdots,2n+1)$ and $\CP_{n+1}$ form the Clifford algebra in $(2n+2)$- dimensions and the representation is known as $2^{n+1}$.

Let us discuss the case of $E=\frac{1}{2}c^2$.
The state with $E=\frac{1}{2}c^2$ satisfies
  $	\hat{Q}_{a}^{A} \varphi_{E=\frac{1}{2}c^2}(x)=0 \ (a=1,2,\cdots,2n+1)
$, which is equivalent to 
\begin{eqnarray}
	&& \frac{d}{dx} \tilde{\varphi}_{E=\frac{1}{2}c^2; \lambda_1, \cdots, \lambda_n, \lambda_{n+1}}(x)=0, \label{3-104}
\end{eqnarray}
where we have performed the singular transformation (\ref{Singular3}).
The solution satisfying (\ref{3-104}) always have only $\lambda_{n+1}=1$ state because the state with $\lambda_{n+1}=-1$ have discontinuity except for the original singularities. 
The formal solutions for the state with $E=\frac{1}{2}c^2$  
are given by
\begin{eqnarray}
	\tilde{\varphi}_{E=\frac{1}{2}c^2 ; \lambda_1, \cdots, \lambda_n, \lambda_{n+1}=+1}
	&=& N_{\lambda_1,\cdots, \lambda_n, \lambda_{n+1}=+1}\left[
		\prod_{k=1}^{n} \frac{1}{2} \left(
			1+\lambda_k \CR_k
		\right) 
	\right]. \label{3-105}
\end{eqnarray}
Let us note that the solution (\ref{3-105}) shows that it takes constant values between the two neighboring singularities. 

Let us study the compatibility of the solutions (\ref{3-105}) with the connection conditions (\ref{A_cc3})-(\ref{A_cc2}).
For $b_1=+1$, all the states with $E=\frac{1}{2}c^2$ satisfy the connection conditions (\ref{A_cc1}) and (\ref{A_cc2}), so that the degeneracy is $2^n$, 
while for $b_1=-1$, all the states do not satisfy the connection conditions (\ref{A_cc1}), (\ref{A_cc2}) and are removed from the functional space.

We can introduce the \lq \lq fermion" number operator as 
\begin{equation}
	(-1)^F \equiv \CP_{n+1}
\end{equation}
for $b_2=0$ and $W'(x)=0$ ($c=0$).
We note that $W'(x)=0$ ensures\footnote{Let us note that when $W''(x)=0$, 
$\CP_{n+1}$ anticommutes with $\hat{Q}_a^A$. 
If we regard $\hat{Q}_a^A$ as a supercharge, $\CP_{n+1}$ can be taken as the \lq \lq fermion" number operator.  
}
\begin{eqnarray}
	\{ Q_a^A, (-1)^F \} =0. \label{F1}
\end{eqnarray}
We call states with $(-1)^F = +1 (-1)$ \lq \lq bosonic" (\lq \lq fermionic") ones.
Under the relation (\ref{F1}), $Q_a^A \varphi_{E;\lambda_1, \cdots, \lambda_n}(x)$  
have the opposite eigenvalues of $(-1)^F$ as $\varphi_{E; \lambda_1, \cdots, \lambda_n}(x)$. 

We discuss the Witten index in this model.
The Witten index of an operator ${\cal O}$ with ${\cal O}^2=1$ is defined by 
\begin{equation}
	\Delta_{W;{\cal O}} \equiv N_+^{E=0} -N_-^{E=0},  
\end{equation}
where $N_{\pm}^{E=0}$ denotes the number of 
the zero energy states with ${\cal O}= \pm 1$, respectively and all the nonzero energy states between ${\cal O}=1$ and ${\cal O}=-1$ are degenerate.

The Witten index of ${\cal P}_{n+1}$ for $W'(x)=0$ $(c=0)$\footnote{In this case, $Q_a^A$ is the same as $\hat{Q}_a^A$.} and $b_2=0$ is given by
\begin{equation}
	\Delta_{W; \CP_{n+1}} = 2^n.
\end{equation}
When $W''(x)=0$,  
the zero energy states do not exist.
However, by shifting the energy, $E \to \hat{E}=E-\frac{1}{2}c^2$, there is a nonzero  Witten index for $\hat{E}=0$ $(E=\frac{1}{2}c^2)$, 
\begin{eqnarray}
	\Delta_{W; \CP_{n+1}} =N_+^{\hat{E}=0}-N_-^{\hat{E}=0}= 2^n .
\end{eqnarray} 
We should make a comment on the special case (\ref{S1}) in which the zero energy states exist.
In this case, the Witten index of $\CG_{\CR_1}\cdots \CG_{\CR_n}$ vanishes because $N_+^{E=0}$ and $N_-^{E=0}$ are the same,  
\begin{eqnarray}
	N_{\CG_{\CR_1}\cdots \CG_{\CR_n}=+1}^{E=0}=N_{\CG_{\CR_1}\cdots \CG_{\CR_n}=-1}^{E=0} =2^{n-1}.
\end{eqnarray}
\\
\\
{\bf 3.3.2 Type B model}
\\\\
The connection conditions compatible with the $N=2n+1$ supercharges $Q_1^B, \cdots, Q_{2n+1}^B$ are given by
\begin{eqnarray}
	&&(I_{2^{n+1}}-U(b_1,b_2)) \Phi_{\varphi}=0, \label{B_CC1}\\
	&& (I_{2^{n+1}}+U(b_1,b_2)) \Phi_{\CD^B \varphi}=0 , \label{B_CC2}
\end{eqnarray}
where
\begin{eqnarray}
	U(b_1,b_2)=b_1 (I_2 \otimes \cdots \otimes I_2 \otimes \sigma_1 ) +b_2 (\vec{e}_{\CR_1} \cdot \vec{\sigma} \otimes \cdots \otimes \vec{e}_{\CR_n} \cdot \vec{\sigma}\otimes \sigma_3 )
\end{eqnarray}
with $b_1, b_2 \in {\bf R}$ and $(b_1)^2+(b_2)^2=1$.
We first note that the $\CG_{\CR_k}$ $(k=1,2,\cdots,n)$ commute with $H^B$ and also with each other. 
Since they are also physical under the connection conditions (\ref{B_CC1}), (\ref{B_CC2}), 
we can introduce simultaneous eigenfunctions,
\begin{eqnarray}
&&H^B \varphi_{E;\lambda_1, \cdots, \lambda_n} (x) =E \varphi_{E;\lambda_1, \cdots, \lambda_n} (x), \\
	&& \CG_{\CR_k}\varphi_{E;\lambda_1, \cdots, \lambda_n} (x)=\lambda_k \varphi_{E;\lambda_1, \cdots, \lambda_n} (x) 
\end{eqnarray}
with $\lambda_k = 1$ or $\lambda_k=-1$ for $k=1,2,\cdots,n$.
The supercharges $Q_a^B\ (a=1,2,\cdots, 2n+1)$ and $\CG_{\CR_k}\ (k=1,2,\cdots,n)$ satisfy the relations 
\begin{equation}
	Q_a^B \CG_{\CR_k} =
	\left\{
		\begin{array}{l}
			-\CG_{\CR_k}Q_a^B \quad {\rm for} \ a=2k-1,2k, \\
			+\CG_{\CR_k}Q_a^B \quad {\rm otherwise}.
		\end{array}
	\right. \label{B_R1}
\end{equation}
Noting that $Q_{2k-1}^B=-i Q_{2k}^B \CG_{\CR_k}$, 
we have the relations 
\begin{eqnarray}
	&&Q_{2k-1}^B \varphi_{E;\lambda_1,\cdots,\lambda_n}(x) = -i \lambda_k Q_{2k}^B \varphi_{E;\lambda_1,\cdots, \lambda_n}(x) \propto \varphi_{E; \lambda_1, \cdots, -\lambda_k, \cdots, \lambda_n}(x), \label{B_R2} \\
	&&Q_{2n+1}^B\varphi_{E;\lambda_1,\cdots, \lambda_n}(x) \propto \varphi_{E;\lambda_1,\cdots, \lambda_n}(x), \label{B_R3}
\end{eqnarray}
then 
the degeneracy of the spectrum is given by $2^n$ for $E>0$.
This result can also be  
obtained from an algebraic point of view, that is, 
for fixed nonzero energy $E$, $Q_a^B /\sqrt{E}$ for $a=1,2,\cdots , 2n+1$ form Clifford algebra in $(2n+1)$-dimensions, and the representation is known as $2^n$.

The above argument cannot apply for $E=0$ states because $Q_a^B \varphi_0=0$.
The equations $Q_a^B\varphi_0(x)=0$ are easily solved and we have formal solutions as
\begin{eqnarray}
	\varphi_{0;\lambda_1, \cdots, \lambda_n}(x)
	=N_{\lambda_1, \cdots, \lambda_n}\left[
		\prod_{k=1}^n \frac{1}{2} \left(1+\lambda_k \CG_{\CR_k}\right)
	\right]
	e^{V(x)},
\end{eqnarray}
where  $N_{\lambda_1, \cdots, \lambda_n}$ denotes normalization constants.
Since the Hamiltonian $H^B$ includes only $V'(x)$ (but not $V(x)$ itself), we have an ambiguity to choose $V(x)$ to satisfy
\begin{equation}
	\CP_i V(x) \CP_i =V(x), \quad  i=1,2,\cdots,n+1. \label{2n+1V}
\end{equation}
Under the singular unitary transformation (\ref{Singular3}), the zero energy states become
\begin{eqnarray}
	\tilde{\varphi}_{0;\lambda_1, \cdots, \lambda_n}(x) 
	\equiv
	\CV \varphi_{0;\lambda_1, \cdots, \lambda_n}(x)
	=N_{\lambda_1, \cdots, \lambda_n}\left[
		\prod_{k=1}^n \frac{1}{2} \left(1+\lambda_k \CR_k\right)
	\right]
	e^{V(x)}. \label{B_zero}
\end{eqnarray}
The zero energy states trivially satisfy the half of the connection conditions (\ref{B_CC2}).
Let us discuss the remaining connection conditions (\ref{B_CC1}), which are given by 
eqs. (\ref{A_cc3})-(\ref{A_cc2}). 

The states (\ref{B_zero}) do not satisfy the connection conditions (\ref{A_cc3}), (\ref{A_cc4}).
On the other hand, the connection conditions (\ref{A_cc1}), (\ref{A_cc2}) are neither satisfied by the zero energy states.
There is, however, an exceptional case
\begin{equation}
	b_1=1.  \label{B_sp1} 
\end{equation}
This is because $V(l_0-\epsilon)=V(l_1+\epsilon)$.

The $\CP_{n+1}$ commutes with $\CG_{\CR_k}$ $(k=1,2,\cdots,n)$, $H^B$ and becomes physical only for $b_2=0$.
An accidental degeneracy occurs in this case. 
We can introduce simultaneous eigenfunctions of the Hamiltonian $H^B$, $\CG_{\CR_k}$ $(k=1,2,\cdots,n)$ and $\CP_{n+1}$ such that
\begin{eqnarray}
	H^B \varphi_{E;\lambda_1,\cdots, \lambda_n, \lambda_{n+1}} (x)&=&E \varphi_{E;\lambda_1,\cdots, \lambda_n, \lambda_{n+1}} (x), \\
	 \CG_{\CR_k} \varphi_{E;\lambda_1,\cdots, \lambda_n, \lambda_{n+1}} (x) &=&\lambda_k \varphi_{E;\lambda_1,\cdots, \lambda_n, \lambda_{n+1}} (x), \\
	 \CP_{n+1} \varphi_{E;\lambda_1,\cdots, \lambda_n, \lambda_{n+1}} (x) &=&\lambda_{n+1} \varphi_{E;\lambda_1,\cdots, \lambda_n, \lambda_{n+1}} (x).
\end{eqnarray}
The $Q_a^B$ and $\CP_{n+1}$ satisfy the relations
\begin{eqnarray}
	Q_a^B \CP_{n+1}
	=
	-\CP_{n+1}Q_a^B, \quad a=1,2,\cdots,2n+1. \label{B_R4}
\end{eqnarray}
Eqs. (\ref{B_R1}), (\ref{B_R2}) and (\ref{B_R4}) imply that 
the degeneracy of the spectrum for $E> 0$ is given by $2^{n+1}$.

The zero energy formal solutions are 
\begin{eqnarray}
	&&\varphi_{0;\lambda_1,\cdots, \lambda_n, \lambda_{n+1}=+1}(x) = N_{\lambda_1, \cdots, \lambda_n}\left [\prod_{k=1}^{n} \frac{1}{2} (1+\lambda_k \CG_{\CR_k}) \right] e^{V(x)}. \label{B_Z}
\end{eqnarray}
Here $\varphi_{0;\lambda_1,\cdots, \lambda_n, \lambda_{n+1}=-1}(x)$ do not exist because the states cannot construct without 
new singularities except for the original singular points $x=l_s\ (s=0,1,\cdots, 2^n-1)$. 
For $b_1=1$, all the zero energy states (\ref{B_Z}) become supersymmetric vacuum states compatible with the connection conditions (\ref{A_cc1}), (\ref{A_cc2}), while, for $b_1=-1$, all the states are incompatible with the connection conditions (\ref{A_cc1}), (\ref{A_cc2}).

The $\CP_{n+1}$ can be regarded as the \lq \lq fermion" number operator only if $b_2=0$, 
\begin{equation}
	(-1)^F =\CP_{n+1}.
\end{equation}
The operator satisfies
\begin{eqnarray}
	&&\{ Q_a^B , (-1)^F \}=0.
\end{eqnarray}

There is a nonzero Witten index of $\CP_{n+1}$ only for $b_1=1$ and $b_2=0$, 
\begin{equation}
	\Delta_{W;\CP_{n+1}} =
	2^n .
\end{equation}
\section{Reduction of the supersymmetry}

\subsection{General discussion}

In this section, we provide a general discussion on a reduction of the supersymmetry by relaxing the connection conditions\footnote{
Falomir and Pisani have also discussed the reduction of supersymmetry in the presence of singular superpotentials\cite{Falomir-Pisani}.}.
So far, we have considered the connection conditions compatible with all the $2n+1$ supercharges.
Here we require that only a subset of the supercharges are physical. To put it the other way around,
we allow some of the $2n+1$ supercharges to become incompatible with connection conditions.

Let us consider the $m$ supercharges that are a subset of the $2n+1$ supercharges (3.1) ((3.2)).
The connection conditions compatible with the $m$ supercharges are given by
\begin{eqnarray}
	&&\left( I_{2^{n+1}} -U \right) \Phi_{\varphi} =0, \\
	&&\left( I_{2^{n+1}} +U \right) \Phi_{\CD^{A(B)} \varphi} =0, 
\end{eqnarray}
where the constraints on the characteristic matrix $U$ are 
\begin{eqnarray}
	U^{\dagger} U &=&I_{2^{n+1}}, \label{4-3}\\
	U^2 &=& I_{2^{n+1}}, \label{4-4}\\
	\gamma_i U \gamma_i &=& -U. \label{4-5} 
\end{eqnarray}
Here the number of $\gamma_i$ is $m$, which is a subset of the $\gamma_a$ $(a=1,2,\cdots,2n+1)$.
The difference from the case of the $N=2n+1$ supersymmetry is in
eq. (4.5), that is, the number of $\gamma_i$ is smaller than that of the case of the $N=2n+1$ supersymmetry.
The smaller the number of $\gamma_i$ in eq. (4.5) becomes, the larger the number of parameters of the characteristic matrix $U$ becomes.
This implies that the supercharges not belonging to the subset are not the physical operators. 
\subsection{\boldmath{$N=2n$} supersymmetry}

In this section, we consider some $N=2n$ supersymmetric models as examples of the reduced supersymmetry. 
We choose $2n$ supercharges that are subset of the $2n+1$ supercharges (\ref{Q_A}), (\ref{Q_B}) as
\begin{eqnarray} 
\begin{array}{ll}
	{\rm (a)} \ Q_{1}^{A},Q_{2}^{A}, \cdots, Q_{2n}^{A}, 
	&\qquad 
	 {\rm (b)}\ Q_{1}^{A},Q_{2}^{A}, \cdots, Q_{2n-1}^{A}, Q_{2n+1}^{A}, \nonumber \\[0.3cm]
	 {\rm (c)}\ Q_{1}^{B},Q_{2}^{B}, \cdots, Q_{2n}^{B},
	 &\qquad 
	 {\rm (d)}\ Q_{1}^{B},Q_{2}^{B}, \cdots, Q_{2n-1}^{B}, Q_{2n+1}^{B}.
\end{array}
\end{eqnarray}
We will obtain the connection conditions compatible with the constraints (\ref{4-3})-(\ref{4-5}) and study the degeneracy of the spectrum and the zero energy states in each model. 
\\
\\
{\bf 4.2.1 (a) $Q_1^A,Q_2^A, \cdots, Q_{2n}^A$}
\\
\\
In order for the supercharges $Q_1^A, \cdots, Q_{2n}^A$ to satisfy the superalgebra, we need only eq. (\ref{3-8}).
And the Hamiltonian is also given by eq. (\ref{3-14}).
The $N=2n$ supersymmetry has already been found and discussed in Ref. \cite{susy4}.
The results of the connection conditions and degeneracy of the spectrum in this subsection agree with those in Ref. \cite{susy4} although the analysis has been incomplete in Ref. \cite{susy4}.

The connection conditions compatible with the $2n$ supercharges $Q_1^A, \cdots, Q_{2n}^A$ are
\begin{eqnarray}
	&&(I_{2^{n+1}}-U)\Phi_{\varphi}=0, \label{a_CC1}\\
	&&(I_{2^{n+1}}+U)\Phi_{\CD^A \varphi}=0, \label{a_CC2}
\end{eqnarray}
where 
\begin{eqnarray}
	U^2&=&I_{2^{n+1}}, \label{a_U1}\\
	 U^{\dagger}U &=&I_{2^{n+1}}, \label{a_U2} \\
	\gamma_i U\gamma_i &=& - U, \qquad  i=1,2,\cdots, 2n. \label{a_U}
\end{eqnarray}
In order to obtain the expression of the characteristic matrix $U$, we use the expansion (\ref{A_U1})-(\ref{A_U3}).
Eq. (\ref{a_U}) implies that the nonvanishing coefficients are only four, 
\begin{eqnarray}
	&&a_1 \equiv C_{\alpha_1=0, \alpha_2=0, \cdots, \alpha_n=0, \alpha_{n+1}=1}, \qquad a_2 \equiv C_{\alpha_1=0, \alpha_2=0, \cdots, \alpha_n=0, \alpha_{n+1}=2}, \\
	&&a_3 \equiv C_{\alpha_1=3, \alpha_2=3, \cdots, \alpha_n=3, \alpha_{n+1}=3}, \qquad  
	a_4 \equiv C_{\alpha_1=3, \alpha_2=3, \cdots, \alpha_n=3, \alpha_{n+1}=0}. 
\end{eqnarray}
Thus, the characteristic matrix $U$ can be expanded as
\begin{eqnarray}
	U &=& a_1  \left( I_2 \otimes \cdots 
	 \otimes I_2 \otimes \sigma_1 \right)
	  +a_2 \left( I_2 \otimes \cdots \otimes I_2 
    	\otimes \sigma_2 \right) \nonumber \\
	&&+a_3 \left( \vec{e}_{{\cal R}_1}\cdot \vec{\sigma} 
    	\otimes \cdots \otimes \vec{e}_{{\cal R}_n}\cdot \vec{\sigma} 
	      \otimes \sigma_3 \right)  
	+a_4 \left( \vec{e}_{{\cal R}_1} \cdot 
	\vec{\sigma}\otimes \cdots \otimes \vec{e}_{{\cal R}_n}
	\cdot \vec{\sigma}\otimes  I_2 \right).
\end{eqnarray}
Let us note the number of parameters in the characteristic matrix $U$ is larger than that in the $N=2n+1$ supersymmetric model because of the decrease of the conditions on the $U$. We further restrict the form of $U$ by imposing  the conditions (\ref{a_U1}) and (\ref{a_U2}). Then, we obtain that 
\begin{equation}
	(a_4)^2=1 \qquad {\rm and}\qquad a_1=a_2=a_3=0, 
\end{equation}
or
\begin{eqnarray}
	(a_1)^2+(a_2)^2+(a_3)^2=1 \qquad {\rm and} \qquad a_4=0 
\end{eqnarray} 
with $a_1, a_2, a_3, a_4 \in {\bf R}$.
Thus, the characteristic matrix $U$ compatible with the $2n$ supersymmetry is given by
\begin{itemize}
\item[(I)] Type I
\end{itemize}
\begin{equation}
	U_{\rm I} (\pm) =\pm \left( \vec{e}_{{\cal R}_1} \cdot 
	\vec{\sigma}\otimes \cdots \otimes \vec{e}_{{\cal R}_n}
	\cdot \vec{\sigma}\otimes  I_2 \right), \label{4-29}
\end{equation}
\begin{itemize}
\item[(II)] Type II
\end{itemize}
\begin{eqnarray}
	U_{\rm II}(a) &=& a_1 \left( I_2 \otimes \cdots 
	 \otimes I_2 \otimes \sigma_1 \right)
	  +a_2 \left( I_2 \otimes \cdots \otimes I_2 
    	\otimes \sigma_2 \right) \nonumber \\
	&&+a_3 \left( \vec{e}_{{\cal R}_1}\cdot \vec{\sigma} 
    	\otimes \cdots \otimes \vec{e}_{{\cal R}_n}\cdot \vec{\sigma} 
	      \otimes \sigma_3 \right) \label{4-30}
\end{eqnarray}
with $(a_1)^2+(a_2)^2+(a_3)^2=1$.

Let us next study the degeneracy of the spectrum. 
We note that $\CG_{\CR_k}$ $(k=1,2,\cdots, n)$ commutes with the Hamiltonian $H^{A}$ and is physical under the connection conditions (\ref{4-29}), (\ref{4-30}), so that 
the state with $E (>0)$ is labeled by the quantum number $\lambda_k$, which is the eigenvalue of $\CG_{\CR_k}$.
Noting that $Q_{2k-1}^A= -i Q_{2k}^A \CG_{\CR_k}$ and the relations
\begin{eqnarray}
	Q_i^A \CG_{\CR_k}=\left\{
		\begin{array}{l}
			-\CG_{\CR_k} Q_i^A \quad {\rm for}\  i=2k-1\ {\rm or}\ 2k, \\
			+\CG_{\CR_k} Q_i^A \quad {\rm otherwise}
		\end{array}
	\right. 
\end{eqnarray}
for $k=1,2,\cdots,n$, we have the relations (\ref{A_2k-1_2k}) for $k=1,2,\cdots, n$.
Thus, the degeneracy for $E> 0$ is given by $2^n$.
This result can also be  
obtained from an algebraic point of view; for fixed nonzero energy $E$, $Q_i^A/\sqrt{E}$ for $i=1,2,\cdots, 2n$ form the Clifford algebra in $2n$-dimensions, and the representation is known as $2^n$.

Let us study the zero energy states.
The states, which satisfy eq. (\ref{3.69}), are formally given by \begin{eqnarray}
 	\varphi_{0; \lambda_1, \cdots, \lambda_n} (x) =
 	N_{\lambda_1, \cdots, \lambda_n} \left[
		\prod_{k=1}^n \frac{1}{2} \left( 1+\lambda_k \CG_{\CR_k} \right)\right]
	e^{-\lambda_1\cdots \lambda_n W(x)} \label{a_Z1}
\end{eqnarray}
with $\lambda_k = 1$ or $-1$ for $k=1,2,\cdots, n$. 
The zero energy states trivially satisfy the half of the connection conditions (\ref{a_CC2}).
We need to verify whether the zero energy states (\ref{a_Z1}) satisfy the remaining connection conditions (\ref{a_CC1}) or not.
In order to investigate the consistency of the zero energy states with the connection conditions, let us perform the singular transformation (\ref{Singular3}), under which the zero energy states become
\begin{eqnarray}
	\tilde{\varphi}_{0; \lambda_1, \cdots, \lambda_n}(x) \equiv \CV \varphi_{0; \lambda_1, \cdots, \lambda_n} (x)
	=
	N_{\lambda_1, \cdots, \lambda_n} \left[
		\prod_{k=1}^n \frac{1}{2} \left( 1+\lambda_k \CR_k \right)\right]
	e^{-\lambda_1\cdots \lambda_n W(x)}. \label{a_Z2}
\end{eqnarray}

According to the singular transformation, the connection conditions (\ref{a_CC1}) can be written, depending on the type I, II,  as 
\begin{itemize}
		\item[(I)] Type I with $U_{{\rm I}}(\pm)$ 
\end{itemize}
\begin{eqnarray}
	&&\left( \CP_1^{\frac{1-\lambda_1}{2}} \cdots\CP_n^{\frac{1-\lambda_n}{2}}\tilde{\varphi}_{\lambda_1,\cdots,\lambda_n}\right)(l_0-\epsilon)=0, \label{a_CC3} \\
	&&\left( \CP_1^{\frac{1-\lambda_1}{2}} \cdots\CP_n^{\frac{1-\lambda_n}{2}}\tilde{\varphi}_{\lambda_1,\cdots,\lambda_n}\right)(l_1+\epsilon)=0 \quad {\rm for} \ 
	\lambda_1\cdots \lambda_n=\mp1 , \label{a_CC6}
\end{eqnarray}
where the double sign of $\lambda_1\cdots \lambda_n$ correspond to the double sign in the $U_{{\rm I}}(\pm)$
\begin{itemize}
		\item[(II)] Type II with $U_{{\rm II}}(a)$ 
\begin{eqnarray}
	\left( \CP_1^{\frac{1-\lambda_1}{2}} \cdots\CP_n^{\frac{1-\lambda_n}{2}}\tilde{\varphi}_{\lambda_1,\cdots,\lambda_n}\right)(l_1+\epsilon)=0 \quad {\rm for} \ 
	a_3\lambda_1\cdots \lambda_n=+1 , \label{a_CC5}
\end{eqnarray}
	\begin{eqnarray}
		&&(1	-a_3 \lambda_1\cdots \lambda_n) \left( \CP_1^{\frac{1-\lambda_1}{2}} \cdots\CP_n^{\frac{1-\lambda_n}{2}}\tilde{\varphi}_{\lambda_1,\cdots,\lambda_n}\right)(l_0-\epsilon)
		\nonumber \\
		&&
		 \qquad \qquad
		-(a_1 -ia_2)  \left( \CP_1^{\frac{1-\lambda_1}{2}} \cdots\CP_n^{\frac{1-\lambda_n}{2}}\tilde{\varphi}_{\lambda_1,\cdots,\lambda_n}\right)(l_1+\epsilon)=0 \nonumber \\
		&&\qquad \qquad \qquad \qquad \qquad \qquad \qquad \qquad 
\quad {\rm for}\ {\rm otherwise}. \label{a_CC4}
	\end{eqnarray}
\end{itemize}
Let us first discuss the case of $U_{\rm I}(+)$. We note that the connection condition (\ref{a_CC1}) does not yield nontrivial conditions for the state with $\lambda_1\cdots\lambda_n=+1$, while the other connection condition (\ref{a_CC2}) is automatically satisfied due to (\ref{3.69}).
On the other hand, we must check whether the state with $\lambda_1\cdots\lambda_n=-1$ satisfies the connection conditions  (\ref{a_CC3}), (\ref{a_CC6}) or not, but it is obviously impossible due to the exponential factor in the solutions (4.20). 
Hence, the degeneracy is given by $2^{n-1}$. 
The discussion goes the same as above for the case of $U_{\rm I}(-)$, so that the degeneracy is, again, $2^{n-1}$.
Thus, the supersymmetry is unbroken.
We have seen that once we determine the connection condition $U_{\rm I}(\pm)$, the states with $\lambda_1\cdots\lambda_n=\pm 1$
can satisfy the connection conditions.

For the type II connection conditions, the zero energy states (\ref{a_Z2}) are found to be inconsistent with the connection conditions (\ref{a_CC5}), (\ref{a_CC4}), so that there are no vacuum states with zero energy. 
There is, however, an exception.
If the following relations are satisfied
\begin{eqnarray}
	\sqrt{\frac{1-a_3}{1+a_3}}=e^{W(l_0-\epsilon)-W(l_1+\epsilon)} , \quad 
	a_1=\sqrt{1-(a_3)^2}, \quad 
	a_2=0, \label{a_sp1}
\end{eqnarray}
all the states (\ref{a_Z2}) accidentally become supersymmetric vacuum states compatible with the connection conditions (\ref{a_CC4}), hence, the degeneracy is $2^n$.
Therefore, for the type II connection conditions, supersymmetry is spontaneously broken except for the above case.

If $a_2=0$, the supercharge $Q_{2n+1}^{A}$ become physical. 
Then, $Q_{1}^{A}, \cdots, Q_{2n+1}^{A}$ form the $N=2n+1$ superalgebra if we require the conditions (\ref{2n+1W}).
Therefore, the supersymmetry is enhanced to the $N=2n+1$ supersymmetry, and the degeneracy has been studied in the previous section.

We can introduce the \lq \lq fermion" number operator as
\begin{equation}
	(-1)^F \equiv \CG_{\CR_1} \cdots \CG_{\CR_n},
\end{equation}
which is physical under the connection conditions (\ref{a_CC1}) and (\ref{a_CC2}), and commutes (anticommutes) with the Hamiltonian $H^A$ (all the supercharges $Q_i^A \ (i=1,2,\cdots ,2n)$).
Let us note that eigenvalues of the fermion number operator are $\lambda_1 \cdots \lambda_n=\pm1 $.
The Witten index of $\CG_{\CR_1}\cdots \CG_{\CR_n}$ with $U_{\rm I}(\pm)$ is given by 
\begin{eqnarray}
	\Delta_{W;\CG_{\CR_1}\cdots \CG_{\CR_n}}=\pm 2^{n-1},  
\end{eqnarray}
where the double sign correspond to the double sign in the $U_{{\rm I}}(\pm)$.

If $W'(x)=0$, the $\CP_{n+1}$ is physical under the connection condition (\ref{4-29}).
And it (anti)commutes with the (supercharges) $H^A$.
Then, we can introduce the Witten index of $\CP_{n+1}$, 
\begin{eqnarray}
	\Delta_{W;\CP_{n+1}}= 2^{n-1}. 
\end{eqnarray}

For the type II connection conditions with $U_{{\rm II}}(a)$, the Witten index of $\CG_{\CR_1}\cdots \CG_{\CR_n}$ vanishes because there are no zero energy states. 
We should make a comment on the special case (\ref{a_sp1}), in which the zero energy states exist. 
The Witten index of $\CG_{\CR_1}\cdots \CG_{\CR_n}$ is, however,  zero  because $N_+^{E=0}$ and $N_-^{E=0}$ are the same,
\begin{eqnarray}	N_{\CG_{\CR_1}\cdots\CG_{\CR_n}=+1}^{E=0}=N_{\CG_{\CR_1}\cdots\CG_{\CR_n}=-1}^{E=0}=2^{n-1}.
\end{eqnarray}
\\\\
{\bf 4.2.2 (b) $Q_{1}^A,Q_2^A, \cdots, Q_{2n-1}^A, Q_{2n+1}^A$}
\\
\\
The $N=2n$ supercharges $Q_{1}^A,Q_2^A, \cdots, Q_{2n-1}^A, Q_{2n+1}^A$ form the $N=2n$ superalgebra if the function $W'(x)$  obeys (3.8) and (\ref{2n+1W}).
The constraints on the characteristic matrix $U$ are 
\begin{eqnarray}
	U^2&=&I_{2^{n+1}}, \label{b_U1}\\
	U^{\dagger}U&=&I_{2^{n+1}}, \label{b_U2}\\
	\gamma_i U\gamma_i &=&-  U, \qquad 
	i=1,2,\cdots, 2n-1,2n+1.\label{b_U3}
\end{eqnarray}
We expand the $U$ in the expression (\ref{A_U1})-(\ref{A_U3}).
Because of the conditions (\ref{b_U3}), nonzero coefficients are only four, 
\begin{eqnarray}
	\begin{array}{ll}
	 a_1 \equiv C_{\alpha_1=0, \alpha_2=0, \cdots, \alpha_n =0, \alpha_{n+1}=1}, &\qquad 
	a_2 \equiv C_{\alpha_1=3, \alpha_2=3, \cdots, \alpha_{n-1} =3, \alpha_n =2, \alpha_{n+1}=3}, \\
	 a_3 \equiv C_{\alpha_1=3, \alpha_2=3, \cdots, \alpha_n =3, \alpha_{n+1}=3}, & \qquad 
	 a_4 \equiv C_{\alpha_1=0, \alpha_2=0, \cdots, \alpha_{n-1} =0,\alpha_n=1, \alpha_{n+1}=1}.
	 	\end{array}
\end{eqnarray}
The conditions (\ref{b_U1}) and (\ref{b_U2}) imply that 
\begin{eqnarray}
	&& (a_4)^2=1 \qquad {\rm and} \qquad a_1=a_2=a_3=0,
\end{eqnarray}
or 
\begin{eqnarray}
	&& (a_1)^2+ (a_2)^2+(a_3)^2=1 \qquad {\rm and}\qquad a_4=0
\end{eqnarray}
with $a_1, a_2, a_3, a_4 \in {\bf R}$.
Thus, the characteristic matrix $U$ compatible with the $2n$ supercharges $Q_1^A, \cdots, Q_{2n-1}^A, Q_{2n+1}^A$ are
\begin{itemize}
\item[(I)] Type I
\end{itemize}
\begin{equation}
	U_{\rm I} (\pm) =\pm \left(
		I_2 \otimes \cdots \otimes I_2 \otimes \vec{e}_{\CP_n}\cdot \vec{\sigma} \otimes \sigma_1
	\right), \label{4-54}
\end{equation}
\begin{itemize}
\item[(II)] Type II
\end{itemize}
\begin{eqnarray}
	U_{\rm II}(a) &=&
	 a_1 \left( I_2 \otimes \cdots 
	 	\otimes I_2 \otimes \sigma_1 \right) \nonumber \\
	 &&
	  +a_2 \left( \vec{e}_{{\cal R}_1} \cdot \vec{\sigma} \otimes \cdots \otimes \vec{e}_{{\cal R}_{n-1}}\cdot \vec{\sigma} \otimes \vec{e}_{{\cal Q}_n}\cdot \vec{\sigma} \otimes \sigma_3 
	  \right) \nonumber \\
	&&+a_3 \left( \vec{e}_{{\cal R}_1}\cdot \vec{\sigma} 
    		\otimes \cdots \otimes \vec{e}_{{\cal R}_n}\cdot \vec{\sigma} 
	      \otimes \sigma_3 \right) \label{4-55}
\end{eqnarray}
with $(a_1)^2+(a_2)^2+(a_3)^2=1$.
With the above characteristic matrix $U$, the connection conditions are given by
\begin{eqnarray}
	&&(I_{2^{n+1}}-U)\Phi_{\varphi}=0, \label{b_CC1}\\
	&&(I_{2^{n+1}}+U)\Phi_{\CD^A \varphi}=0. \label{b_CC2}
\end{eqnarray}

In this case, $\CG_{\CR_n}$ is no longer physical under the connection conditions (\ref{4-54}), (\ref{4-55}), but
instead, $\CG_{\CP_n} \CP_{n+1}$ is physical.
The Hamiltonian $H^A$, $\CG_{\CR_k}$ $(k=1,2,\cdots,n-1)$ and $\CG_{\CP_n} \CP_{n+1}$ are physical, and they commute with each other.
We can introduce simultaneous eigenfunctions of these operators 
such that
\begin{eqnarray}
	H^A \varphi_{E;\lambda_1, \cdots, \lambda'_n} (x) 
		&=&E \varphi_{E;\lambda_1, \cdots, \lambda'_n} (x), \\
	\CG_{\CR_k} \varphi_{E;\lambda_1, \cdots, \lambda'_n} (x) 
		&=&\lambda_k  \varphi_{E;\lambda_1, \cdots, \lambda'_n} (x), \quad k=1,2,\cdots, n-1,\\
	\left( \CG_{\CP_n} \CP_{n+1} \right)\varphi_{E;\lambda_1, \cdots, \lambda'_n} (x) 
		&=&\lambda'_n  \varphi_{E;\lambda_1, \cdots, \lambda'_n} (x)
\end{eqnarray}
with $\lambda_k=1$ or $-1$ for $k=1,2,\cdots,n-1$ and $\lambda'_{n}= 1$ or $-1$.
In this case, we have the relations
\begin{eqnarray}
	&& Q_{2k-1}^A \varphi_{E; \lambda_1,\cdots, \lambda'_n}(x) =-i \lambda_k Q_{2k}^A \varphi_{E; \lambda_1,\cdots, \lambda'_n}(x)  \propto \varphi_{E; \lambda_1,\cdots, -\lambda_k, \cdots, \lambda'_n}(x), \nonumber \\
	&& \qquad \qquad \qquad \qquad \qquad \qquad \qquad \qquad \qquad  
	k=1,2,\cdots,n-1, \\
	&&Q_{2n-1}^A \varphi_{E; \lambda_1,\cdots, \lambda'_n}(x) \propto 
	\varphi_{E; \lambda_1,\cdots, -\lambda'_n}(x), \\
	&&Q_{2n+1}^A \varphi_{E; \lambda_1,\cdots, \lambda'_n}(x) \propto 
	\varphi_{E; \lambda_1,\cdots, -\lambda'_n}(x),
\end{eqnarray}
because
\begin{eqnarray}
	\CG_{\CR_k} Q_i^A &=&
		\left\{
			\begin{array}{l}
				-Q_i^A \CG_{\CR_k} \quad {\rm for} \ k=2i-1,2i, \\
				+Q_i^A \CG_{\CR_k} \quad {\rm otherwise},
			\end{array}
		\right.  
		\\
	\left( \CG_{\CP_n} \CP_{n+1}\right) Q_i^A&=&- Q_i^A \left( \CG_{\CP_n} \CP_{n+1} \right)  \quad 
\end{eqnarray}
for $i=1,2,\cdots,2n-1,2n+1$ and $Q_{2k-1}^A =-i Q_{2k}^A\CG_{\CR_k}$ for $k=1,2,\cdots,n-1$. 
Thus, the degeneracy for $E> 0$ is given by $2^n$.
This result can also be  
obtained from an algebraic point of view; for fixed nonzero energy $E$, $Q_i^A/\sqrt{E}$ for $i=1,2,\cdots, 2n-1,2n+1$ form the Clifford algebra in $2n$-dimensions, and the representation is known as $2^n$.

The above argument cannot apply for the zero energy states.
The zero energy states are given by solving   
\begin{eqnarray}
	\CD^A \varphi_0(x)=\left[ \left(\CR_1 \cdots \CR_{n+1} \frac{d}{dx}\right) + \lambda_1 \cdots \lambda_{n-1} \CG_{\CR_n} \left(\CR_1\cdots\CR_{n+1} W'(x)\right) \right] \varphi_{0; \lambda_1,\cdots, \lambda'_n}(x)=0. \nonumber \\ \label{4.66}
\end{eqnarray}
The solutions to eq. (\ref{4.66}) are formally given by
\begin{eqnarray}
	\varphi_{0;\lambda_1,\cdots, \lambda'_n }(x)
	&=& N_{\lambda_1, \cdots, \lambda'_n} \left(
		\prod_{k=1}^{n-1} \frac{1+\lambda_k \CG_{\CR_k}}{2}
	\right) \nonumber \\
	&&
	\times \left\{
		\left(\frac{1+\CG_{\CR_n}}{2}\right)
			e^{-\lambda_1\cdots \lambda_{n-1} W(x)}
		+\lambda'_n 
			\left(\frac{1-\CG_{\CR_n}}{2}\right)
				e^{\lambda_1 \cdots \lambda_{n-1} W(x)}
	\right\}, 
\end{eqnarray}
and trivially satisfy the half of the connection conditions (\ref{b_CC2}).
We need to verify whether the zero energy states satisfy the remaining connection conditions (\ref{b_CC1}) or not.
Under the singular unitary transformation (\ref{Singular3}), the zero energy states become
\begin{eqnarray}
	\tilde{\varphi}_{0;\lambda_1,\cdots, \lambda'_n }(x)
	&\equiv&
	\CV\varphi_{0;\lambda_1,\cdots, \lambda'_n }(x) \nonumber \\
	&=& N_{\lambda_1, \cdots, \lambda'_n} \left(
		\prod_{k=1}^{n-1} \frac{1+\lambda_k \CR_k}{2}
	\right) \nonumber \\
	&&
	\times \left\{
		\left(\frac{1+\CR_n}{2}\right)
			e^{-\lambda_1\cdots \lambda_{n-1} W(x)}
		+\lambda'_n 
			\left(\frac{1-\CR_n}{2}\right)
				e^{\lambda_1 \cdots \lambda_{n-1} W(x)}
	\right\}, \label{4.68}
\end{eqnarray}
and accordingly the connection conditions, (\ref{b_CC1}) 
 can be written as
\begin{itemize}
	\item [(I)] Type I with $U_{{\rm I}}(\pm)$
\end{itemize}
\begin{eqnarray}
	&&\left( \CP_1^{\frac{1-\lambda_1}{2}} \cdots\CP_{n-1}^{\frac{1-\lambda_{n-1}}{2}}\tilde{\varphi}_{\lambda_1,\cdots,\lambda'_n}\right)(l_0-\epsilon)=0, \label{b_CC3} \\
	&&\left( \CP_1^{\frac{1-\lambda_1}{2}} \cdots\CP_{n-1}^{\frac{1-\lambda_{n-1}}{2}}\tilde{\varphi}_{\lambda_1,\cdots,\lambda'_n}\right)(l_1+\epsilon)=0 \qquad \qquad {\rm for} \ 
	\lambda'_n=\mp1, \label{b_CC6}
\end{eqnarray}
where the double sign of $\lambda'_n$ correspond to the double sign in the $U_{{\rm I}}(\pm)$,
\begin{itemize}
	\item[(II)] Type II with $U_{{\rm II}}(a)$
\end{itemize}
\begin{eqnarray}
	\left( \CP_1^{\frac{1-\lambda_1}{2}} \cdots\CP_{n-1}^{\frac{1-\lambda_{n-1}}{2}}\tilde{\varphi}_{\lambda_1,\cdots,\lambda'_{n}}\right)(l_1+\epsilon)=0 \qquad {\rm for}\ a_3 \lambda_1\cdots \lambda_{n-1}=+1,  \label{b_CC5}
\end{eqnarray}
\begin{eqnarray}
	&&(1-a_3\lambda_1\cdots \lambda_{n-1})\left( \CP_1^{\frac{1-\lambda_1}{2}} \cdots\CP_{n-1}^{\frac{1-\lambda_{n-1}}{2}}\tilde{\varphi}_{\lambda_1,\cdots,\lambda'_n}\right)(l_0-\epsilon) \nonumber \\
	&&\qquad 
	-(a_1 -i a_2 \lambda_1\cdots\lambda'_n) \left( \CP_1^{\frac{1-\lambda_1}{2}} \cdots\CP_{n-1}^{\frac{1-\lambda_{n-1}}{2}}\tilde{\varphi}_{\lambda_1,\cdots,\lambda'_n}\right)(l_1+\epsilon)=0  \nonumber \\
	&&\qquad \qquad \qquad \qquad \qquad \qquad \qquad \qquad \qquad \qquad \qquad {\rm for}\ {\rm otherwise}. \label{b_CC4}
\end{eqnarray} 

We first consider the case of $U_{\rm I}(+)\ \left( U_{\rm I}(-) \right)$.
The zero energy states with $\lambda'_n=+1 \ \left( \lambda'_n=-1\right)$ can satisfy the connection conditions (\ref{b_CC1}), (\ref{b_CC2}) because the connection conditions (\ref{b_CC1}) do not yield nontrivial conditions for the states with $\lambda'_n=+1\ \left( \lambda'_n=-1\right)$, and the connection conditions (\ref{b_CC2}) are automatically satisfied due to  (\ref{4.66}).
On the other hand, the states with $\lambda'_n=-1 \ \left( \lambda'_n=+1 \right)$ do not satisfy the connection conditions (\ref{b_CC3}), (\ref{b_CC6}) because of the exponential factor in the solution.
Thus, the degeneracy of the zero energy states for the case of $U_{\rm I}(\pm)$ is given by $2^{n-1}$, and the supersymmetry is unbroken.

For the case of $U_{\rm II}(a)$, 
the zero energy states do not satisfy the connection conditions 
(\ref{b_CC5}), (\ref{b_CC4}) with one exception. 
If the following relations are satisfied
\begin{eqnarray}
	\sqrt{\frac{1-a_3}{1+a_3}}=e^{2 W(l_0-\epsilon)},\quad 
	a_1 = \sqrt{1-(a_3)^2}, \quad 
	a_2=0,  \label{b_sp1}
\end{eqnarray}
all the states (\ref{4.68}) are compatible with the  
connection conditions (\ref{b_CC4}), then, the degeneracy is $2^n$. 
Hence, for the type II connection conditions, the supersymmetry is spontaneously broken except for the above case. 

Note that if $a_2=0$, the supercharge $Q_{2n}^{A}$ becomes physical. 
Then, the supersymmetry is enhanced to the $N=2n+1$ supersymmetry.

In this model, $\CG_{\CP_{n}}\CP_{n+1}$ can be regarded as the \lq \lq fermion" number operator, 
\begin{eqnarray}
	(-1)^F = \CG_{\CP_n}\CP_{n+1}.
\end{eqnarray}
The Witten indices of $\CG_{\CP_n}\CP_{n+1}$, $\CP_{n+1}$ for $U_{{\rm I}}(\pm)$ are given by
\begin{eqnarray}
	\Delta_{W;\CG_{\CP_n}\CP_{n+1}}&=&\pm 2^{n-1}, \label{4.2.2:1}
	\\
	\Delta_{W;\CP_{n+1}}&=&2^{n-1}, \quad \qquad 
	{\rm only\ for}\ W'(x)=0,
\end{eqnarray}
where the double sign in eq. (\ref{4.2.2:1}) correspond to the double sign in the $U_{{\rm I}}(\pm)$.
For $U_{{\rm II}}(a)$, the Witten index of $\CG_{\CP_n} \CP_{n+1}$ vanishes because the zero energy states do not exist.
Let us comment on the special case (\ref{b_sp1}), where we have the zero energy states.
We note that the supersymmetry is enhanced to the same $N=2n+1$ supersymmetry in the special case (\ref{b_sp1}) as that in the previous section.
\\\\
{\bf 4.2.3 (c) $Q_{1}^{B}, \cdots, Q_{2n}^{B} $}
\\
\\
The $N=2n$ supercharges $Q_{1}^{B}, \cdots, Q_{2n}^{B} $ form the $N=2n$ superalgebra if the $V'(x)$ obeys (\ref{3-10}).
In this case, the constraints on the characteristic matrix $U$ are same with the case of {\bf (a)}.
The connection conditions are given by 
\begin{eqnarray}
	&&(I_{2^{n+1}}-U)\Phi_{\varphi}=0, \label{c_CC1}\\
	&&(I_{2^{n+1}}+U)\Phi_{\CD^B \varphi}=0,  \label{c_CC2}
\end{eqnarray}
where $U$ is  given by  (\ref{4-29}) or (\ref{4-30}).

For $U_{{\rm I}}(\pm)$, the $\CG_{\CR_k}$ $(k=1,2,\cdots,n)$ and $\CP_{n+1}$ are physical under the connection conditions (\ref{c_CC1}),(\ref{c_CC2}) and commute with $H^B$. 
We can introduce simultaneous eigenfunctions of $H^B$, $\CG_{\CR_k}$ $(k=1,2,\cdots,n)$ and $\CP_{n+1}$ such that
\begin{eqnarray}
	H^B \varphi_{E;\lambda_1,\cdots,\lambda_{n+1}}(x) &=&E\varphi_{E;\lambda_1,\cdots,\lambda_{n+1}}(x), \\
	\CG_{\CR_k}\varphi_{E;\lambda_1,\cdots,\lambda_{n+1}}(x)&=&\lambda_k \varphi_{E;\lambda_1,\cdots,\lambda_{n+1}}(x), \qquad 
	k=1,2,\cdots,n, \\
	\CP_{n+1} \varphi_{E;\lambda_1,\cdots,\lambda_{n+1}}(x)&=&\lambda_{n+1}\varphi_{E;\lambda_1,\cdots,\lambda_{n+1}}(x).
\end{eqnarray}
Although the labels in $\varphi$ increase, the degeneracy of the spectrum for $E>0$ states does not increase in this model. 
The $Q_i^B$ $(i=1,2,\cdots ,2n)$, $\CG_{\CR_k}$ $(k=1,2,\cdots,n)$ and $\CP_{n+1}$ satisfy
\begin{eqnarray}
	&& \CP_{n+1} Q_i^B=-Q_i^B \CP_{n+1} \qquad {\rm for} \ i=1,2,\cdots,2n, \\
	&& \CG_{\CR_k}Q_i^B =
	\left\{
		\begin{array}{l}
			-Q_i^B \CG_{\CR_k} \qquad {\rm for} \ i=2k-1,2k,\\
			+Q_i^B \CG_{\CR_k} \qquad {\rm otherwise},
		\end{array}
	\right. 		
\end{eqnarray}
The supercharge $Q_i^B$ $(i=1,2,\cdots,n)$ changes not only the sign of $\lambda_i$ but also the sign of $\lambda_{n+1}$, so that 
no supercharge connects $\lambda_1\cdots \lambda_{n+1}=+1$ states with $\lambda_1\cdots \lambda_{n+1}=-1$ ones.
Therefore, 
The degeneracy between the states with $\lambda_1\cdots\lambda_{n+1}=+1$ and the states with $\lambda_1\cdots\lambda_{n+1}=-1$ is not guaranteed, so that the $2^{n+1}$-fold degeneracy is not ensured.

Since $Q_{2k-1}^B = -i Q_{2k}^B \CG_{\CR_k}$ for $k=1,2,\cdots,n$ and 
\begin{eqnarray}
	&&Q_{2k-1}^B \varphi_{E;\lambda_1, \cdots, \lambda_n}(x)  =-i \lambda_k Q_{2k}^B \varphi_{E;\lambda_1, \cdots, \lambda_n}(x) \propto \varphi_{E;\lambda_1, \cdots, -\lambda_k, \cdots, -\lambda_n}(x) 
\end{eqnarray}
for $k=1,2,\cdots,n$, 
the $2^n$-fold degeneracy for $E>0$ states is ensured. 
This result can also be 
obtained from an algebraic point of view; for fixed nonzero energy $E$, $Q_i^B/\sqrt{E}$ for $i=1,2,\cdots, 2n$ form the Clifford algebra in $2n$-dimensions, and the representation is known as $2^n$.
The $\CP_{n+1}$ is regarded as a \lq \lq fermion" number operator $(-1)^F$ in the type I connection conditions.
For $U_{{\rm II}}(a)$, since the $\CP_{n+1}$, in general,  is not physical, the degeneracy for $E>0$ is $2^n$.

The zero energy states satisfying $Q_i^B \varphi_0 (x)=0$ are formally
\begin{eqnarray}
	\varphi_{0;\lambda_1, \cdots, \lambda_n}(x)=N_{\lambda_1, \cdots, \lambda_n} \left(\prod_{k=1}^{n} \frac{1+\lambda_k \CG_{\CR_k}}{2}\right) e^{V(x)}.
\end{eqnarray}
The zero energy states trivially satisfy the half of the connection conditions (\ref{c_CC2}).
Under the singular unitary transformation (\ref{Singular3}), the remaining connection conditions (\ref{c_CC1}) can be written by eqs. (\ref{a_CC3})-(\ref{a_CC4}), and the zero energy states become
\begin{eqnarray}
	\tilde{\varphi}_{0;\lambda_1, \cdots, \lambda_n}(x)\equiv \CV \varphi_{0;\lambda_1, \cdots, \lambda_n}(x)=N_{\lambda_1, \cdots, \lambda_n} \left(\prod_{k=1}^{n} \frac{1+\lambda_k \CR_k}{2}\right) e^{V(x)}. \label{4.2.3:1}
\end{eqnarray}

Let us first discuss the case of $U_{\rm I}(+)\ (U_{\rm I}(-))$.
We note that the connection conditions (\ref{c_CC1}) do not yield nontrivial conditions for the states with $\lambda_1\cdots \lambda_n=+1\ (\lambda_1\cdots \lambda_n=-1)$, while the other connection conditions (\ref{c_CC2}) are automatically satisfied due to $\CD^B \varphi_0 (x)=0$.
On the other hand, we must check whether the states with $\lambda_1 \cdots \lambda_n=-1\ (\lambda_1\cdots \lambda_n=+1)$ satisfy the connection conditions (\ref{a_CC3}), (\ref{a_CC6}) or not. 
It is obviously impossible because of the exponential factor in the solution.
Hence the degeneracy for the zero energy states for the case of $U_{\rm I}(\pm)$ is given by $2^{n-1}$, and the supersymmetry is unbroken.
%
We have seen that once we determine the connection conditions $U_{\rm I}(\pm)$, the state with $\lambda_1\cdots \lambda_n=\pm 1$ can satisfy the connection conditions.  

For the type II connection conditions, the zero energy states (\ref{4.2.3:1}) are found to be inconsistent with the connection conditions 
(\ref{a_CC5}), (\ref{a_CC4}), so that there are no vacuum states with zero energy.
There is, however, an exception.
If the following relation is satisfied
\begin{eqnarray}
	a_1=1,
\end{eqnarray}
all the states (\ref{4.2.3:1}) accidentally become supersymmetric vacuum states compatible with the connection conditions (\ref{a_CC4}), then, the degeneracy is $2^n$. 
Hence, for the type II connection conditions, supersymmetry is spontaneously broken except for the above case.
Note that if $a_2=0$, the supercharge $Q_{2n+1}^{B}$ becomes physical, therefore the supersymmetry is enhanced to the $N=2n+1$ supersymmetry, which is same as the type B model in the previous section. 

The fermion number operators can be defined as 
\begin{eqnarray}
	&&(-1)^F=\CG_{\CR_1}\cdots \CG_{\CR_n} \qquad {\rm for} \ U_{{\rm I}}(\pm) \ {\rm or} \ U_{{\rm II}}(a), \\
	&&(-1)^F =\CP_{n+1} \qquad \qquad \quad {\rm for} \ U_{{\rm I}}(\pm) \ {\rm or} \ U_{{\rm II}}(a_1=\pm 1).
\end{eqnarray}
For $U_{{\rm I}}(\pm)$, the Witten indices of $\CG_{\CR_1}\cdots\CG_{\CR_n}$ and $\CP_{n+1}$ are given by 
\begin{eqnarray}
	&&\Delta_{W; \CG_{\CR_1}\cdots \CG_{\CR_n}}=\pm 2^{n-1}, \label{4.2.3:4-72}\\ 
	&&\Delta_{W;\CP_{n+1}}=2^{n-1}, 
\end{eqnarray}
where the double sign in eq. (\ref{4.2.3:4-72}) correspond to the double sign in the $U_{{\rm I}}(\pm)$.
Let us comment on the special case of $U_{\rm II}(a)$ with $a_1=1$, where we have the zero energy states.
We note that the supersymmetry is enhanced to the same $N=2n+1$ supersymmetry even in the special case of $a_1=1$ as that in the previous section.
\\\\
{\bf 4.2.4 (d) $Q_{1}^{B}, \cdots, Q_{2n-1}^{B}, Q_{2n+1}^{B}$}
\\
\\
The $N=2n$ supercharges $Q_{1}^{B}, \cdots, Q_{2n-1}^{B}, Q_{2n+1}^{B}$ form the $N=2n$ superalgebra if the $V'(x)$ are assumed to obey (\ref{3-10}).
The characteristic matrix $U$ is again given by (\ref{4-54}),(\ref{4-55}).
The connection conditions become
\begin{eqnarray}
	&& (I_{2^{n+1}}-U)\Phi_{\varphi}=0, \label{d_CC1}\\
	&& (I_{2^{n+1}}+U)\Phi_{\CD^B \varphi}=0. \label{d_CC2}
\end{eqnarray}
In this model, $\CG_{\CR_k}$ $(k=1,2,\cdots,n-1)$ and  $\CG_{\CP_n} \CP_{n+1}$ are physical and commute with $H^B$. 
Since these operators commute with each other,
we can introduce simultaneous eigenfunctions of $H^B$, $\CG_{\CR_k}$ $(k=1,2,\cdots,n-1)$ and $\CG_{\CP_n} \CP_{n+1}$ such that 
\begin{eqnarray}
	H^B \varphi_{E;\lambda_1, \cdots, \lambda'_n} (x) 
		&=&E \varphi_{E;\lambda_1, \cdots, \lambda'_n} (x), \\
	\CG_{\CR_k} \varphi_{E;\lambda_1, \cdots, \lambda'_n} (x) 
		&=&\lambda_k  \varphi_{E;\lambda_1, \cdots, \lambda'_n} (x), \quad k=1,2,\cdots, n-1,\\
	\left( \CG_{\CP_n} \CP_{n+1} \right)\varphi_{E;\lambda_1, \cdots, \lambda'_n} (x) 
		&=&\lambda'_n  \varphi_{E;\lambda_1, \cdots, \lambda'_n} (x).
\end{eqnarray}
Since
\begin{eqnarray}
	\CG_{\CR_k}Q_i^B &=&
		\left\{	
			\begin{array}{l}
				-Q_i^B \CG_{\CR_k} \qquad {\rm for} \ k=2i-1,2i, \\
				+Q_i^B \CG_{\CR_k} \qquad {\rm otherwise},
			\end{array}
		\right. \\
	\left(\CG_{\CP_n}\CP_{n+1}\right) Q_i^B &=&-Q_i^B\left( \CG_{\CP_{n}}\CP_{n+1}\right)
\end{eqnarray}
and $Q_{2k-1}^B =-i Q_{2k}^B\CG_{\CR_k}$ for $k=1,2,\cdots, n-1$, 
we have the relations 
\begin{eqnarray}
	&&Q_{2k-1}^B \varphi_{E;\lambda_1, \cdots, \lambda'_n}(x)  =-i \lambda_k Q_{2k}^B \varphi_{E;\lambda_1, \cdots, \lambda'_n}(x) \propto \varphi_{E;\lambda_1, \cdots, -\lambda_k, \cdots, -\lambda'_n}(x), \nonumber \\
	&&\qquad \qquad \qquad \qquad \qquad \qquad \qquad \qquad \qquad \quad 
	k=1,2,\cdots, n-1, \\
	&&Q_{2n-1}^B \varphi_{E;\lambda_1, \cdots, \lambda'_n}(x) \propto \varphi_{E;\lambda_1, \cdots,  -\lambda'_n}(x),\\
	&&Q_{2n+1}^B \varphi_{E;\lambda_1, \cdots, \lambda'_n}(x) \propto \varphi_{E;\lambda_1, \cdots,  -\lambda'_n}(x).
\end{eqnarray}
Then, the degeneracy of the spectrum for $E> 0$ is given by $2^n$.
This result can also be 
obtained from an algebraic point of view; for fixed nonzero energy $E$, $Q_i^B/\sqrt{E}$ for $i=1,2,\cdots, 2n-1,2n+1$ form the Clifford algebra in $2n$-dimensions, and the representation is known as $2^n$.

The zero energy states are given by solving 
\begin{eqnarray}
	\CD^B \varphi_0(x)= \left[ \left(\CR_1 \cdots \CR_{n+1}\frac{d}{dx} \right) -\CP_{n+1} \left( \CR_1 \cdots \CR_{n+1} V'(x) \right) \right] \varphi_{0; \lambda_1, \cdots, \lambda'_n}(x)=0.
\end{eqnarray}
The solutions are formally written by
\begin{eqnarray}
	\varphi_{0;\lambda_1,\cdots, \lambda'_n}(x)=N_{\lambda_1,\cdots, \lambda'_n}\left(\prod_{k=1}^{n-1} \frac{1+\lambda_k \CG_{\CR_k}}{2} \right) 
	\left\{
	\left(\frac{1+\CG_{\CR_n}}{2}\right)+\lambda'_n \left(\frac{1-\CG_{\CR_n}}{2}\right)\right\} e^{V(x)}. \label{d_Z1}
\end{eqnarray}
The zero energy states trivially satisfy the half of the connection conditions (\ref{d_CC2}). 
Under the singular unitary transformation (\ref{Singular3}), the remaining connection conditions (\ref{d_CC1}) become eqs. (\ref{b_CC3})-(\ref{b_CC4}), and accordingly, the zero energy states (\ref{d_Z1}) become
\begin{eqnarray}
	\tilde{\varphi}_{0;\lambda_1,\cdots, \lambda'_n}(x)&\equiv& \CV \varphi_{0;\lambda_1\cdots \lambda'_n}(x) \nonumber \\
	&=& N_{\lambda_1,\cdots, \lambda'_n}\left(\prod_{k=1}^{n-1} \frac{1+\lambda_k \CR_k}{2} \right) 
	\left\{
	\left(\frac{1+\CR_n}{2}\right)+\lambda'_n \left(\frac{1-\CR_n}{2}\right)\right\} e^{V(x)}. \nonumber \label{4.2.4:1}\\
\end{eqnarray}

We first consider the case of $U_{\rm I}(+)\ \left( U_{\rm I}(-) \right)$.
The zero energy states with $\lambda'_n=+1 \ \left( \lambda'_n=-1\right)$ can satisfy the connection conditions (\ref{d_CC1}), (\ref{d_CC2}) because the connection conditions (\ref{d_CC1}) do not yield nontrivial conditions for the states with $\lambda'_n=+1\ \left( \lambda'_n=-1\right)$, and the connection conditions (\ref{d_CC2}) are automatically satisfied due to  ${\cal D}^{B} \varphi_0(x)=0$. 
On the other hand, the states with $\lambda'_n=-1 \ \left( \lambda'_n=+1 \right)$ do not satisfy the connection conditions (\ref{b_CC3}), (\ref{b_CC6}). 
Hence, the degeneracy of the zero energy states for the case is given by $2^{n-1}$, and the supersymmetry is unbroken.

For the case of $U_{\rm II}(a)$, 
the zero energy states (\ref{4.2.4:1}) do not satisfy the connection conditions 
(\ref{b_CC5}), (\ref{b_CC4}) with one exception. 
If 
\begin{eqnarray}
	a_1=1, \label{d_sp1}
\end{eqnarray}
all the states (\ref{4.2.4:1}) accidentally become supersymmetric vacuum states compatible with the connection conditions (\ref{b_CC4}), then, the degeneracy is $2^n$. 
Thus, for the type II connection conditions, the supersymmetry is spontaneously broken except for the above case.

The fermion number operator can be introduced as
\begin{eqnarray}
	(-1)^F = \CG_{\CP_n} \CP_{n+1}.
\end{eqnarray}
For $U_{{\rm I}}(\pm)$, the Witten indices of $\CG_{\CP_{n}}\CP_{n+1}$ and $\CP_{n+1}$ are given by
\begin{eqnarray}
	&&\Delta_{W;\CG_{\CP_n}\CP_{n+1}}=\pm 2^{n-1}, \label{4.2.4:4-89}\\
	&&\Delta_{W; \CP_{n+1}}=2^{n-1}, 
\end{eqnarray}
where the double sign in eq. (\ref{4.2.4:4-89}) correspond to the double sign in the $U_{{\rm I}}(\pm)$.
For $U_{{\rm II}}(a)$, the Witten index of $\CG_{\CP_n}\CP_{n+1}$ vanishes because the zero energy states do not exist.
Let us comment on the special case (\ref{d_sp1}), where we have the zero energy states. We note that the supersymmetry is enhanced to the same $N=2n+1$ supersymmetry in the special case as that in the previous section.

\section{Summary}

In this paper, we have studied the supersymmetry on a circle with $2^n$ point singularities placed at regular intervals in a full detail. 
The two types of the $N=2n+1$ supercharges (type A and type B) are constructed in terms of the $n+1$ sets of discrete transformations $\{\CP_k, \CQ_k,\CR_k\}$ $(k=1,2,\cdots,n+1)$. 
The singularities can be described by the connection conditions in our formulation.
We have found the connection conditions make all the supercharges physical, so that the $N=2n+1$ supersymmetry can be realized under the connection conditions.

In our analysis, the connection conditions, under which arbitrary subset of the $N=2n+1$ supercharges are physical and others are not physical, can be obtained.
Thus, the $N=2n+1$ supersymmetry can be reduced to $M$-extended supersymmetry for any integer $M<N$ due to the connection conditions.

We have also studied the degeneracy of the spectrum in particular zero energy states in the $N=2n+1$ supersymmetric models and some $N=2n$ supersymmetric models as examples of the reduction of the supersymmetry. 
The energy eigenfunctions can be labeled by the eigenvalues of the physical operators which commute with the Hamiltonian and each other.
The degeneracy of the spectrum for the states with $E>0$ have been  obtained by discussing the transition of the labeled states under the action of the supercharges.
The analysis cannot apply for the states with $E=0$.
This is because the zero energy states satisfy $Q \varphi_0 =0$, that is, the transition does not occurs.
The zero energy states, however, satisfy the first order differential equation thanks to the supersymmetry, and 
the formal solutions can be obtained.
We have checked whether the formal solutions satisfy the connection conditions or not.
The degeneracy of the spectrum and the Witten index are summarized in table 1 and table 2.

%
\begin{table}[h]
\begin{center}
\caption{degeneracy of the spectrum in the $N=2n+1$ supersymmetric models}
	\begin{tabular}{|c|c|c|l|c|}
		\hline
		Type & parameters & \multicolumn{2}{|c|}{degeneracy}& Witten index  \\
		\hline
		Type A & $\begin{array}{c} b_2=0 \\ {\rm and} \end{array}$  
		& $E>\frac{1}{2}c^2$ & $2^{n+1}$ &
		\\
		\cline{3-4} &
		$\begin{array}{c} W''(x)=0 \\ (W'(x)\equiv c) \end{array} $& $E=\frac{1}{2}c^2$ & $\left\{ \begin{array}{l} 2^n \quad {\rm for }\ b_1=+1 \\ 0 \ \ \ \ {\rm for}\ b_1=-1\end{array}\right. $ & 
		$\begin{array}{c} \Delta_{W;\CP_{n+1}}=2^n \\
		 ({\rm for}\  b_1=1) 
		 \end{array}$ \\
		\cline{2-5} & otherwise & $E>0$&$2^n$&
		\\
		 \cline{3-4} &&$E=0$ & $\left\{\begin{array}{l} 2^n\quad {\rm for}\ * 
		 \\  0\quad {\rm for}\ {\rm otherwise} \end{array}\right.$ &0 
		 \\
		 \hline
		 Type B & $b_2=0$  & $E>0$ & $2^{n+1}$&\\
		\cline{3-4} && $E=0$ & $\left\{ \begin{array}{l} 2^n \quad {\rm for }\ b_1=+1 \\ 0 \quad {\rm for}\ b_1=-1\end{array}\right.$ &$\begin{array}{c}
		\Delta_{W; \CP_{n+1}}=2^n \\
		({\rm for}\ b_1=1)
	\end{array}$ \\
		\cline{2-5} & otherwise & $E>0$&$2^n$ & \\
		 \cline{3-4} &&$E=0$ & 0&0  \\
		 \hline
		 \multicolumn{4}{l}{where $*$ denotes $ \sqrt{\frac{1-b_2}{1+b_2}}=e^{2W(l_0-\epsilon)}.$}
	\end{tabular}
\end{center}
\end{table}
\begin{table}[h]
\begin{center}
\caption{degeneracy of the spectrum in the $N=2n$ supersymmetric models}
	\begin{tabular}{|c|c|c|c|l|}
		\hline
		   Type  & $U_{\rm I}$ or $U_{{\rm II}}$ & \multicolumn{2}{|c|}{degeneracy} & Witten index \\
		  \hline
		  $a$ & $ U_{\rm I}(\pm)$ & $E>0$ & $2^n$& $\Delta_{W;\CG_{\CR_1}\cdots\CG_{\CR_n}}=\pm 2^{n-1}$
		   \\
		  \cline{3-4} && $E=0$ & $2^{n-1}$&$\Delta_{W;\CP_{n+1}} = 2^{n-1}$ for $W'(x)=0$ \\
		  \cline{2-5} & $U_{{\rm II}}(a)$ & $E>0$ & $2^n$&  
		  \\
		  \cline{3-4} && $E=0$ & $
		  \left\{\begin{array}{l}  2^n \quad {\rm for}\ *
\\ 0\quad {\rm for}\ {\rm otherwise} \end{array}\right.$ & 
0
\\
		\hline
	  $b$ & $ U_{\rm I}(\pm)$ & $E>0$ & $2^n$& $\Delta_{W;\CG_{\CP_n} \CP_{n+1}}=\pm 2^{n-1}$ \\
		  \cline{3-4} && $E=0$ & $2^{n-1}$& $\Delta_{W;\CP_{n+1}}=2^{n-1}\ {\rm for}\ W'(x)=0$ \\
		  \cline{2-5} & $U_{{\rm II}}(a)$ & $E>0$ & $2^n$ & 		  \\
		  \cline{3-4} && $E=0$ & $ 
		  0
		   $ & 0
\\
		\hline
	  $c$ & $ U_{\rm I}(\pm)$ & $E>0$ & $2^n$ &
	  $\Delta_{W; \CG_{\CR_1}\cdots\CG_{\CR_n}}=\pm 2^{n-1}$ 
	  \\
		  \cline{3-4} && $E=0$ & $2^{n-1}$ &$\Delta_{W;\CP_{n+1}}=2^{n-1}$\\
		  \cline{2-5} & $U_{{\rm II}}(a)$ & $E>0$ & $2^n$ &
		  \\
		  \cline{3-4} && $E=0$ & $
		  0$ &
		 0
\\
		\hline
		 $d$ & $ U_{\rm I}(\pm)$ & $E>0$ & $2^n$&  $\Delta_{W; \CG_{\CP_n}\CP_{n+1}}=\pm 2^{n-1}$
		 \\
		  \cline{3-4} && $E=0$ & $2^{n-1}$& $\Delta_{W; \CP_{n+1}}= 2^{n-1}$
		  \\
		  \cline{2-5} & $U_{{\rm II}}(a)$ & $E>0$ & $2^n$ &
		  \\
		  \cline{3-4} && $E=0$ & $
		  0
		 $&0 
\\
		\hline 
		\multicolumn{5}{l}{where $*$ denotes $ 
			\sqrt{\frac{1-a_3}{1+a_3}}=e^{W(l_0-\epsilon)-W(l_1+\epsilon)}, a_1= \sqrt{1-(a_3)^2}, a_2=0, $} 	\end{tabular}
\end{center}
\end{table} 
\section*{Acknowledgements}
The authors wish to thank
C.S. Lim,
I. Tsutsui
for valuable discussions and useful comments. 
M. S. is supported in part by the Grant-in-Aid for
Scientific Research (No.15540277) by the Japanese Ministry
of Education, Science, Sports and Culture.
K. T. is supported in part by 
the Grant-in-Aid for Science Research, Ministry of Education, Science
and Culture, Japan, No. 17043007 and the $21$st Century 
COE Program at Osaka University.

\end{document}